\documentclass{article}

\usepackage{PRIMEarxiv}
\usepackage[utf8]{inputenc}
\usepackage[T1]{fontenc}
\usepackage{hyperref}
\usepackage{url}
\usepackage{booktabs}
\usepackage{amsfonts}
\usepackage{nicefrac}
\usepackage{microtype}
\usepackage{lipsum}
\usepackage{fancyhdr}
\usepackage{graphicx}
\graphicspath{{media/}}
\usepackage{amsmath,amssymb,mathtools,amsthm,xcolor}
\usepackage{caption,subcaption}
\usepackage{array}
\usepackage{tabularx}
\usepackage{lmodern}
\usepackage{enumitem}
\usepackage{multirow}
\usepackage{algorithm}
\usepackage{algorithmic}
\usepackage{wrapfig}
\usepackage{setspace}
\usepackage{pgfplots}
\usepackage{tikz}
\usetikzlibrary{shapes,arrows.meta,positioning,fit,calc,decorations.pathreplacing}
\pgfplotsset{compat=1.18}

\newcommand{\R}{\mathbb{R}}
\newcommand{\bz}{\mathbf{z}}
\newcommand{\bp}{\mathbf{p}}
\newcommand{\bx}{\mathbf{x}}

\newcommand{\br}{\mathbf{r}}
\newcommand{\bmu}{\boldsymbol{\mu}}

\newcommand{\cN}{\mathcal{N}}

\title{
PhyloSDF: Phylogenetically-Regularized Neural Generation of 3D Skull Morphology via Residual Flow Matching
}

\author{
  Kaikwan Lau \\
  Department of Mathematics \\
  The Chinese University of Hong Kong \\
  \texttt{kaikwan.lau@link.cuhk.edu.hk} \\
   \And
  Gary P.\,T.\ Choi \\
  Department of Mathematics \\
  The Chinese University of Hong Kong \\
  \texttt{ptchoi@cuhk.edu.hk} \\
}

\begin{document}
\maketitle

\begin{abstract}
Generating complete 3D skulls from a few scans per species requires a model that shares information across species while retaining their morphological differences. We present PhyloSDF, which integrates a phylogenetically regularized DeepSDF representation with Residual Conditional Flow Matching (Residual CFM). Each latent code is decomposed into an empirical species centroid and a residual; a shared flow learns residual variation pooled across species. We evaluate the model on 100 micro-CT skulls of Darwin's Finches and relatives across 24 species, with 2--7 scans per species. Among samplers sharing a fixed tree-trained representation, Residual CFM achieves the lowest mean species-mixture MMD-CD over three seeds ($1.175\times10^{-3}$, versus $1.198\times10^{-3}$ for empirical residual sampling and $1.408\times10^{-3}$ for a per-species Gaussian mixture). Empirical residual sampling performs better in one seed. Complementary proximity diagnostics place one of 300 outputs within the decoded-training-copy distance range. As a biological check, a separate proxy-based morphometric audit with automated cranial measurements finds agreement in species-level proportions (median correlation 0.98), and generated skulls follow an orbit-curvature rule not used in training (median error 5.8\%, against 5.3\% for raw scans). Dated-tree neighbours improve centroid prediction for withheld species over the global mean; fitted codes allow extension when scans become available. In four focal species, generated skulls retain 47--61\% of real within-species mesh diversity. Altogether, PhyloSDF offers a practical way to generate and explore 3D skull morphology when specimens are scarce.
\end{abstract}


\section{Introduction} \label{sec:intro}

Three-dimensional morphological data {support} comparative biology, paleontology, and evolutionary developmental biology~\cite{cunningham2014virtual, davies2017open, klingenberg2010evolution}. {Micro-computed tomography ($\mu$CT) digitizes specimens as dense surface meshes~\cite{keklikoglou2019micro} for quantitative morphometric analysis~\cite{bardua2019practical}. Yet studies often contain only tens to a few hundred specimens because museum material is limited~\cite{mulqueeney2024many} and scanning, segmentation and quality control are costly.}

{Learning a generative distribution from so few specimens is difficult. Variational autoencoders~\cite{kingma2014auto}, generative adversarial networks~\cite{goodfellow2020generative}, diffusion models~\cite{ho2020denoising, lau2026efficient}, and flow matching~\cite{lipman2022flow} have been successful on images and 3D shapes, often with much larger training sets. At this scale, overfitting and limited coverage are concerns~\cite{van2021memorization,zhang2021understanding}, but their severity depends on the model and training protocol. Biological morphology also has a useful source of structure: species share evolutionary history~\cite{felsenstein1985phylogenies}. Phylogenetic relationships can therefore guide how information is shared across species.}

\begin{figure*}[h!]
\centering
\includegraphics[width=1\linewidth]{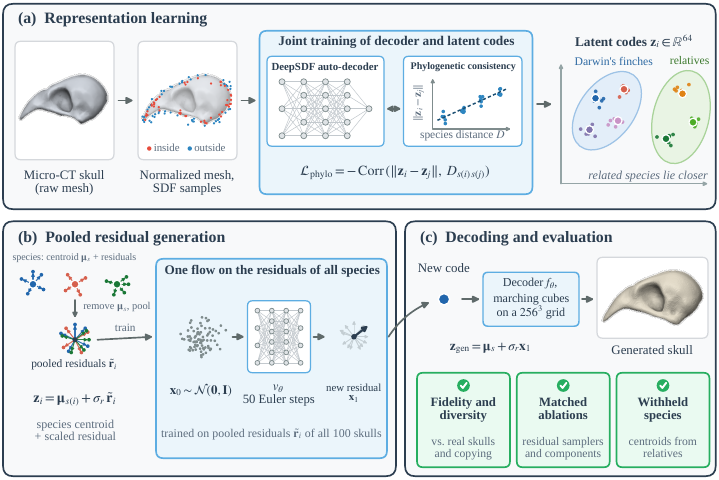}
\caption{\textbf{Overview of the proposed pipeline for phylogenetically-{regularized} 3D skull generation.} (a)~Data preprocessing and representation. Raw STL skull meshes {are} converted to watertight meshes and normalized, after which signed distance function (SDF) samples are generated for DeepSDF auto-decoder training {with a Phylogenetic Consistency Loss}. (b) Generative modeling in latent space. A Residual Conditional Flow Matching (CFM) network{, trained on the residuals of all species,} learns to generate intra-species residuals, deviations of individual latent codes from their species centroid. (c) Generation and evaluation. New latent codes are decoded by the frozen DeepSDF decoder into full 3D skull meshes {and evaluated}.}
\label{FIG_introduction}
\end{figure*}

{In this work, we ask whether a neural model can generate complete 3D skulls with measured species-level proportions from only a few scans per species. Three challenges guide the design: estimating within-species variation in a 64-dimensional latent space, incorporating species relationships, and distinguishing generated variation from proximity to training specimens. Note that image-based FID~\cite{heusel2017gans} is unsuitable for these meshes, and small reference sets make distributional evaluation difficult~\cite{chong2020effectively}. Therefore, we combine mesh distances, morphometry, and complementary generativity diagnostics. Specifically,} we introduce PhyloSDF, a phylogenetically {regularized} neural generative model for 3D biological morphology that addresses {these} challenges. Figure~\ref{FIG_introduction} illustrates the complete PhyloSDF pipeline, which maps raw micro-CT meshes into a phylogenetically structured latent space, factorizes generative modeling via Residual Flow Matching, and {evaluates the generated meshes with mesh-level and morphometric measures}.

{First, a \textbf{Phylogenetic Consistency Loss} regularizes DeepSDF~\cite{park2019deepsdf} so that latent distances correlate with supplied species distances. LPC reaches about 0.99 without worsening reconstruction in our experiments. This measures fit to the supervising matrix; shuffled-matrix controls and a retrained species holdout assess what that fit contributes.}

{Second, \textbf{Residual Conditional Flow Matching} separates empirical species-centroid lookup from a learned distribution of residual variation pooled across species. The centroid-plus-residual decomposition follows hierarchical random-effects models~\cite{laird1982random,gelman2007data}; our contribution combines it with an implicit shape representation and a residual flow. Pooling lets specimens from different species inform a shared within-species distribution, without creating additional independent observations.}

{Finally, we evaluate representation fit with LPC, generated meshes with fidelity, coverage and diversity measures, and species-level proportions with automated morphometry. MorphFID summarizes distributions of interpretable geometric descriptors. A leave-one-species-out centroid analysis is complemented by a retrained holdout that predicts species absent from representation learning.}

{Our primary objective is generation from sparse species samples; the secondary objectives are to isolate the learned sampler's contribution and assess the role of species relationships. This setting reflects collections that grow a few specimens at a time and include species with no scans.}

\section{Related Work} \label{sec:related}

\subsection{Neural Implicit 3D Representations}

{Implicit neural representations encode surfaces as continuous fields.} DeepSDF \cite{park2019deepsdf} introduced the auto-decoder architecture in which a shared neural network maps a latent code and a 3D query point to a signed distance value, enabling smooth surface reconstruction at arbitrary resolution and supporting direct latent-space interpolation without an encoder network. Occupancy Networks~\cite{mescheder2019occupancy} proposed an alternative binary occupancy formulation that {learns a continuous occupancy classifier} while preserving resolution-independence. IM-NET~\cite{chen2019learning} demonstrated that implicit decoders can be conditioned on global shape embeddings to generate diverse shapes from compact codes. Subsequent work on Neural Unsigned Distance Fields~\cite{chibane2020neural} and {Differentiable Volumetric Rendering}~\cite{niemeyer2020differentiable} extended these ideas to open surfaces and appearance modeling, respectively. {Dinnage and Kleineberg~\cite{dinnage2025generative} trained DeepSDF on the bills of about 2,000 bird species and related its latent space to ecology, a study of representation across many species rather than generation within species from a few specimens.} PhyloSDF builds directly on the DeepSDF auto-decoder architecture, extending it with a phylogenetic consistency loss that structures the latent space geometrically according to evolutionary relationships, a constraint that prior methods do not consider.

\subsection{Generative Models for 3D Shapes}

Generative modeling of 3D geometry has been approached across a spectrum of representations. Point-cloud methods use {latent adversarial models and Gaussian mixtures~\cite{achlioptas2018learning}} or normalizing flows{~\cite{yang2019pointflow}} to model unstructured point sets{, and Li et al.~\cite{li2023progressive} generate point clouds progressively with consistency across hierarchical levels}, while voxel-based approaches~\cite{maturana2015voxnet} apply 3D convolutional architectures to occupancy grids but suffer from cubic memory scaling. Mesh-based generative models~\cite{nash2020polygen} have demonstrated sequential generation of {the vertices and faces of} triangle meshes. For neural implicit representations specifically, DualSDF~\cite{hao2020dualsdf} introduced generative frameworks operating in the latent spaces of implicit decoders. More recent work has applied DDPM~\cite{ho2020denoising, lau2026efficient} to 3D shape latents~\cite{vahdat2022lion} {and to signed-distance fields~\cite{shim2023diffusion}}, achieving high-quality results on large-scale datasets such as ShapeNet with tens of thousands of training examples. {Yang et al.~\cite{yang2026unified} supply several types of conditions as inputs to one latent diffusion model, whereas PhyloSDF supplies the species only through its centroid.} Flow matching~\cite{lipman2022flow} has emerged as a more training-stable alternative{, used by CellFlux~\cite{zhang2025cellflux} to model cellular morphology under perturbation}.

{Many of these methods were developed and evaluated on thousands of shapes. Their performance on roughly 100 specimens cannot be assumed from those results. Limited-data generation often adapts a pretrained generator~\cite{abdollahzadeh2025survey} or augments training data, including point clouds~\cite{zhu2024advancements}; in our latent setting, augmentation reduced mesh fidelity (Section~\ref{sec:ablation_results}). PhyloSDF instead pools residual variation across species.}

\subsection{Phylogenetic Comparative Methods and Machine Learning}

Incorporating phylogenetic information into machine learning has been explored in several related contexts. {Phylo-NN~\cite{elhamod2023discovering} organizes image embeddings by phylogenetic level, Phylo-Diffusion~\cite{khurana2024hierarchical} conditions image generation on the tree of life, and G2PDiffusion~\cite{liu2025g2pdiffusion} generates morphological images from DNA alignments and environmental inputs; these methods work with images and large training sets rather than complete 3D shapes from a few specimens.} Pagel's $\lambda$~\cite{pagel1999inferring} and related statistics quantify the degree to which trait covariance matches the expectation from a known phylogeny, but these methods operate on scalar or low-dimensional trait vectors rather than on neural latent representations. To our knowledge, no prior work has used phylogenetic patristic distances as a regularizer on the geometry of a neural {3D} shape representation. The preceding survey reveals a gap: no existing method jointly addresses extreme data scarcity, phylogenetic structure, and 3D morphological generation.

\section{Mathematical Background}
\label{sec:background}

\subsection{Implicit Surface Representation}
\label{sec:bg_sdf}

Let $\Omega \subset \mathbb{R}^3$ be a compact region with $C^1$ boundary $\partial \Omega$. The signed distance function (SDF)~\cite{gropp2020implicit} encodes the geometry of $\partial\Omega$ as a scalar field:
\begin{equation}
    \phi_\Omega(\mathbf{p})
    \;=\;
    \operatorname{sgn}(\mathbf{p})\;
    \inf_{\mathbf{q}\,\in\,\partial\Omega}
    \|\mathbf{p} - \mathbf{q}\|_2,
    \label{eq:sdf}
\end{equation}
where $\operatorname{sgn}(\mathbf{p}) = -1$ {if} $\mathbf{p} \in \Omega$ {and $+1$ otherwise}, so that $\partial \Omega$ is recovered as the zero level-set $\{\mathbf{p} \in \mathbb{R}^3 : \phi_\Omega(\mathbf{p}) = 0\}$. A key property is $\|\nabla\phi_\Omega\|_2 = 1$ almost everywhere (the Eikonal condition), which {characterizes ideal} SDFs~\cite{osher2004level}{; it is not a training penalty in this study}.

DeepSDF~\cite{park2019deepsdf} parameterizes the mapping $\phi_\Omega$ with a neural network $f_\theta : \mathbb{R}^D \times \mathbb{R}^3 \to \mathbb{R}$, taking a latent code $\mathbf{z} \in \mathbb{R}^D$ and a query point $\mathbf{p} \in \mathbb{R}^3$ as input. In the auto-decoder formulation, each training specimen $i$ owns a dedicated learnable code $\mathbf{z}_i$, optimized jointly with the shared decoder weights $\theta$ by minimizing
\begin{equation}
    \displaystyle     \mathcal{L}_{\mathrm{rec}}(\theta, {\mathbf{z}_i})
    =
    \frac{1}{|\mathcal{P}_i|}
\sum_{\mathbf{p} \in \mathcal{P}_i}
    \Big|
      \operatorname{clamp}_\delta\!\bigl(f_\theta(\mathbf{z}_i, \mathbf{p})\bigr)
      -
      \operatorname{clamp}_\delta\!\bigl(\phi_i(\mathbf{p})\bigr)
    \Big|,
    \label{eq:recon}
\end{equation}
where $\phi_i(\mathbf{p})$ is the ground-truth signed distance for specimen $i$, $\operatorname{clamp}_\delta(x) = \max(-\delta, \min(\delta, x))$ with $\delta = 0.1$ suppresses the influence of far-field points, and $\mathcal{P}_i$ is a {fixed random subset} of $5 \times 10^4$ {of $10^5$ precomputed} query points per specimen{, half near the surface and half uniform in $[-1,1]^3$ (SI Section~S1)}.

\subsection{Phylogenetic Distance}
\label{sec:bg_phylo}

A phylogenetic tree $\mathcal{T} = (V, E, \ell)$ is a rooted, edge-weighted tree in which leaves correspond to extant taxa, internal nodes correspond to hypothetical common ancestors, and $\ell : E \to \mathbb{R}_{>0}$ assigns evolutionary time (or molecular divergence) to each branch. We define the {patristic distance} $d_{\mathrm{phylo}}(a, b)$ between taxa $a$, $b$ as the total branch length along the unique path connecting them in $\mathcal{T}$:
\begin{equation}
    d_{\mathrm{phylo}}(a, b)
    \;=\;
    \sum_{e \,\in\, \mathrm{path}(a,b)} \ell(e).
    \label{eq:patristic}
\end{equation}
All distances are normalized to $[0, 1]$ {(see Section~\ref{sec:dataset})}.

\subsection{Flow Matching}
\label{sec:bg_fm}

Flow matching~\cite{lipman2022flow,albergo2022building} learns a generative model as a continuous normalizing flow defined by a learned velocity field. Let $p_0 = \mathcal{N}(\mathbf{0}, \mathbf{I}_D)$ be a tractable source distribution and $p_1$ be the target data distribution over {$\mathbb{R}^D$}. The goal is to learn a time-dependent velocity field $v_\theta : \mathbb{R}^D \times [0,1] \to \mathbb{R}^D$ such that integrating the ODE
\begin{equation}
    \frac{d\mathbf{x}_t}{dt}
    \;=\;
    v_\theta(\mathbf{x}_t,\, t),
    \qquad
    \mathbf{x}_0 \sim p_0,
    \label{eq:flow_ode}
\end{equation}
from $t = 0$ to $t = 1$, transports $p_0$ to $p_1$. {Because} the marginal velocity is not available in closed form{,} the conditional flow matching (CFM) objective~\cite{lipman2022flow} {conditions} on individual data points. Given $\mathbf{x}_0 \sim p_0$ and target sample $\mathbf{x}_1 \sim p_1$, sampled independently, we define the straight-line conditional path $\mathbf{x}_t = (1-t)\,\mathbf{x}_0 + t\,\mathbf{x}_1$ with conditional target velocity $u_t(\mathbf{x}_t \mid \mathbf{x}_1) = \mathbf{x}_1 - \mathbf{x}_0$. The CFM training objective is then defined as
\begin{equation}
    \mathcal{L}_{\mathrm{CFM}}
    \;=\;
    {\frac{1}{D}}\,
    \mathbb{E}_{\,t \sim \mathcal{U}[0,1],\;
                 \mathbf{x}_0 \sim p_0,\;
                 \mathbf{x}_1 \sim p_1}
    \Bigl\|
        v_\theta(\mathbf{x}_t,\, t)
        \;-\;
        (\mathbf{x}_1 - \mathbf{x}_0)
    \Bigr\|_2^2,
    \label{eq:cfm_loss}
\end{equation}
{where $1/D$ matches the implemented mean squared error; $\mathcal{L}_{\mathrm{CFM}}$ and the intractable marginal objective have the same gradients in $\theta$~\cite{lipman2022flow}.} Eq.~\eqref{eq:flow_ode} is solved by $K = 50$ Euler steps with step size $\Delta t = 1/K$. The straight-line interpolant uses independent coupling between $p_0$ and $p_1$; this is distinct from OT-CFM~\cite{tong2023improving}, which minimizes transport cost by pairing source and target samples{; independent coupling avoids estimating such a coupling from about 100 samples}.

\subsection{Fr\'{e}chet Distance between Gaussian Distributions}
\label{sec:bg_frechet}

Let $P_k = \mathcal{N}(\boldsymbol{\mu}_k, \boldsymbol{\Sigma}_k)$, $k = 1, 2$, be two distributions on $\mathbb{R}^d$. The squared Fr\'{e}chet distance, equivalently, the squared 2-Wasserstein distance between Gaussians~\cite{dowson1982frechet} admits the closed-form expression
\begin{equation}
    d_F^2(P_1, P_2)
    =
    \|\boldsymbol{\mu}_1 - \boldsymbol{\mu}_2\|_2^2
    +
    \operatorname{tr}\!\Bigl(
        \boldsymbol{\Sigma}_1
        + \boldsymbol{\Sigma}_2
        - 2\,\bigl(
            \boldsymbol{\Sigma}_1^{1/2}
            \boldsymbol{\Sigma}_2
            \boldsymbol{\Sigma}_1^{1/2}
          \bigr)^{\!1/2}
    \Bigr),
    \label{eq:frechet}
\end{equation}
where $(\cdot)^{1/2}$ denotes the unique symmetric positive semidefinite matrix square root. {FID~\cite{heusel2017gans} applies Eq.~\eqref{eq:frechet} to Inception network activations, which are not defined on 3D skull meshes, and with $N = 100$ specimens and a feature dimension $d \sim 2{,}048$ the sample covariance is severely rank-deficient.} Our MorphFID metric (Section~\ref{sec:mfid}) {instead uses} low-dimensional interpretable morphometric descriptors.

\subsection{Chamfer Distance}
\label{sec:bg_cd}
For point clouds $P,Q\subset\R^3$, the Chamfer distance is given by
\begin{equation}
  \displaystyle \mathrm{CD}(P,Q)=\frac{1}{|P|}  \sum_{\bp\in P}\min_{\mathbf{q}\in Q}\|\bp-\mathbf{q}\|_2^2 +\frac{1}{|Q|}   \sum_{\mathbf{q}\in Q}\min_{\bp\in P}\|\mathbf{q}-\bp\|_2^2.
  \label{eq:cd}
\end{equation}

\section{Method} \label{sec:method}

\begin{figure*}[t!]
\centering
\includegraphics[width=1\linewidth]{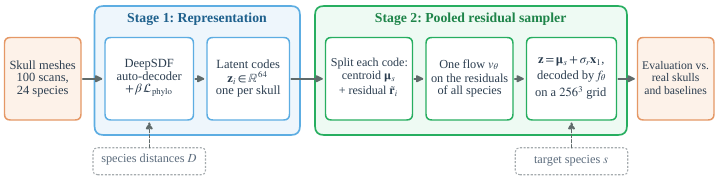}
\caption{\textbf{Overview of our PhyloSDF pipeline.} {Stage~1 fits one code per skull with a shared DeepSDF decoder and a species-distance regularizer. Stage~2 splits each code into its species centroid and a residual and trains one Residual CFM on the residuals of all species. To generate a skull, a sampled residual is added to the target centroid, the code is decoded and Marching Cubes extracts the surface, which is evaluated against real skulls and baselines.}}
\label{fig:overview}
\end{figure*}

PhyloSDF comprises {two} stages (Fig.~\ref{fig:overview}): (1)~Phylogenetically-regularized DeepSDF encoding {and (2)}~Residual CFM training and generation. { Latent augmentation is evaluated separately in Section~\ref{sec:ablation_results}.}

\subsection{Stage 1: Phylogenetically-Regularized DeepSDF} \label{sec:stage1}

We adopt the DeepSDF auto-decoder formulation, in which each of the $N = 100$ specimens owns a learnable latent code $\mathbf{z}_i \in \mathbb{R}^{64}$ optimized jointly with the shared decoder $f_\theta$ via the reconstruction objective $\mathcal{L}_\text{rec}$ defined in Eq.~\eqref{eq:recon}. {We add a \textbf{Phylogenetic Consistency Loss} to align latent geometry with supplied species distances.}

The key observation is that under Brownian motion trait evolution, the expected squared morphological divergence between two taxa is proportional to their patristic distance $d_\text{phylo}(a, b)$~{\cite{felsenstein1985phylogenies}}. {This motivates our regularizer, which maximizes the Pearson correlation between pairwise latent code distances and the supplied species distances; it uses unsquared Euclidean distances and so does not directly enforce the Brownian relationship.} Let $\mathcal{S}$ be {the set of all $N(N-1)/2$ specimen pairs $(i, j)$,} and $d^L_{ij} = \|\mathbf{z}_i - \mathbf{z}_j\|_2$ and $d^P_{ij} = d_\text{phylo}({s(i), s(j)})$ denote the latent and phylogenetic distances respectively{, where $s(i)$ is the species of specimen $i$}. The loss is:
\begin{equation}
  \mathcal{L}_\text{phylo} = -\,\rho\!\left(\{d^L_{ij}\}_{(i,j)\in\mathcal{S}},\;
  \{d^P_{ij}\}_{(i,j)\in\mathcal{S}}\right),
  \label{eq:lphylo}
\end{equation}
where $\rho(\cdot,\cdot)$ is the Pearson correlation coefficient. Minimizing $\mathcal{L}_\text{phylo}$ is equivalent to maximizing the linear alignment between morphological and phylogenetic pairwise distances, making the loss differentiable and directly interpretable via the same LPC metric used at evaluation time{, which therefore measures fit to the supervising matrix}. The full DeepSDF objective is:
\begin{equation}
  \mathcal{L}_\text{DeepSDF} = \mathcal{L}_\text{rec} + \alpha\|\mathbf{z}_i\|_2^2
  + \beta\,\mathcal{L}_\text{phylo},
  \label{eq:ldeepsdf}
\end{equation}
with $\alpha = 10^{-3}$ and $\beta = {5 \times 10^{-3}}$. The phylogenetic loss is activated every 10 batches to reduce computational overhead; between activations, only $\mathcal{L}_\text{rec}$ and the $\ell_2$ regularizer are used. Figure~\ref{sfig:phylo} illustrates the phylogenetic structure that drives $\mathcal{L}_\text{phylo}$. Closely related genera occupy nearby latent regions; distantly related outgroup species are pushed far from the Darwin's Finches clade.

\subsection{Stage {2}: Residual Conditional Flow Matching} \label{sec:stage2}

When {a flow is conditioned on species inputs}, each species provides few target points in $\mathbb{R}^{64}${, so species-specific variation is difficult to estimate (Section~\ref{sec:ablation_results})}. Our solution exploits the group structure of the data. We decompose each latent code into its species centroid and a residual:
\begin{equation}
  \mathbf{z}_i = \boldsymbol{\mu}_{s(i)} + \mathbf{r}_i, \qquad
  \boldsymbol{\mu}_s = \frac{1}{|S_s|}\sum_{j \in S_s} \mathbf{z}_j,
  \label{eq:decomp}
\end{equation}
where $S_s$ is the set of specimens belonging to species $s$, $\mathbf{r}_i$ is the intra-species residual of specimen $i$, and $\boldsymbol{\mu}_s$ is the species centroid{, which is} computed analytically from real codes only and requires no learning. {This is the group-mean decomposition of hierarchical random-effects models~\cite{laird1982random,gelman2007data}, here with empirical centroids and no shrinkage. We train the CFM of Eq.~\eqref{eq:cfm_loss}} to generate normalized residuals $\tilde{\mathbf{r}}_i = (\mathbf{r}_i - \boldsymbol{\mu}_r) / \sigma_r$, where $\boldsymbol{\mu}_r$ and $\sigma_r$ are the {mean and scalar standard deviation of the 100 real residuals ($\boldsymbol{\mu}_r=\mathbf{0}$ by construction)}.

{Within each species, residuals have zero empirical mean. Pooling them allows all 100 specimens to inform one distribution rather than estimating a separate distribution from each species' few scans. It shares information without adding independent observations; centring also constrains residuals within species. The working assumption is a similar form of within-species variation across species. Species, size and phylogeny inputs to the velocity network are held at neutral values, so species identity enters generation only through the centroid.}

{To penalize an estimated variance shortfall during training, we augment} $\mathcal{L}_\text{CFM}$ with a variance-preservation penalty:
\begin{equation}
  \mathcal{L}_\text{var} = \max\!\left(0,\;\mathrm{Var}(\mathbf{x}_1) -
  \mathrm{Var}(\hat{\mathbf{x}}_1)\right),
  \label{eq:lvar}
\end{equation}
where $\hat{\mathbf{x}}_1 = \mathbf{x}_{0.5} + 0.5\,v_\theta(\mathbf{x}_{0.5}, 0.5)$ is a one-step estimate of the generated output evaluated at the trajectory midpoint{, and for a minibatch $X\in\mathbb{R}^{B\times 64}$, $\operatorname{Var}(X)=\frac{1}{64B-1}\sum_{b,d}(X_{bd}-\overline X)^2$ is a single scalar over all entries, with $\overline X$ their mean}. The total training objective is $\mathcal{L} = \mathcal{L}_\text{CFM} + \lambda_v\,\mathcal{L}_\text{var}$, with $\lambda_v = 0.1$. {Because $\mathbf{x}_{0.5}$ contains the training target, the penalty does not control the variance of samples drawn from noise, and removing it made no detectable difference (Section~\ref{sec:ablation_results}). The flow is trained for 10,000 updates on batches of 100 residuals with AdamW (SI Section~S1).}

\subsection{Generation Procedure} \label{sec:generation}
Given a target species $s$ with centroid $\boldsymbol{\mu}_s$, new skull meshes are generated as follows: (i) sample $\mathbf{x}_0 \sim \mathcal{N}(\mathbf{0}, {T^2}\mathbf{I}_{64})$ {with temperature $T=1$}; (ii) integrate the learned velocity field from $t = 0$ to $t = 1$ using $K = 50$ Euler steps with step size $\Delta t = 1/K$; (iii) denormalize the output residual $\hat{\mathbf{r}} = \mathbf{x}_1 \cdot \sigma_r + \boldsymbol{\mu}_r$; (iv) reconstruct the full latent code $\mathbf{z}_\text{new} = \boldsymbol{\mu}_s + \hat{\mathbf{r}}$; and (v) decode $\mathbf{z}_\text{new}$ via Marching Cubes on a $256^3$ voxel grid {over $[-1,1]^3$, keeping open or multi-component surfaces without repair} (see SI {Algorithm~S1}). The velocity network {of Stage~2} is further referred to as \textbf{PhyloFlowNet}{; training hyperparameters, optimizer configurations and network architectures are given in SI Section~S1 and Table~S1}.

\section{Experiments} \label{sec:experiments}

\subsection{Dataset} \label{sec:dataset}

\begin{figure}[h!]
\centering
\includegraphics[width=\linewidth]{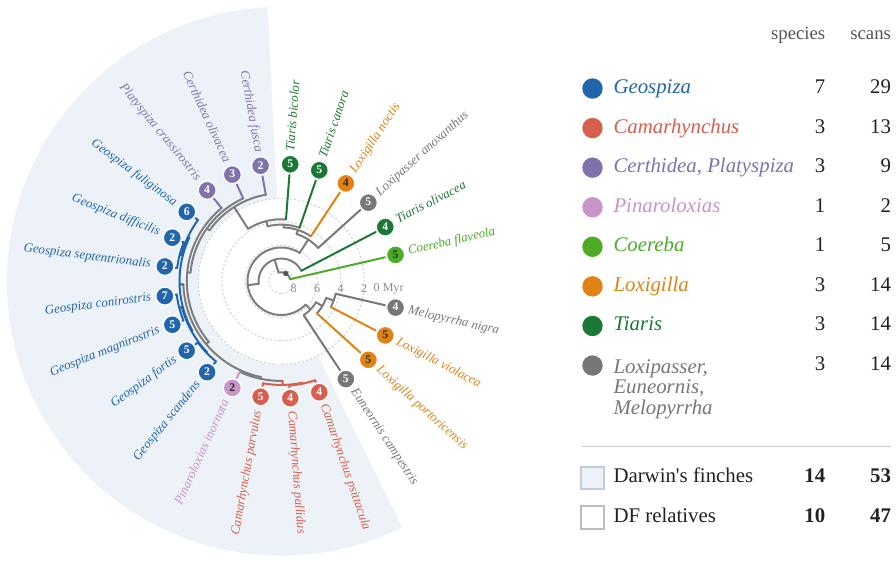}
\caption{\textbf{Phylogenetic tree and dataset composition for the $24$ species and $N = 100$ specimens used in all experiments.} Species are organized into eight biologically defined clades{, with the numbers of species and scans at right}. {Branch lengths come from the dated summary tree of~\cite{mctavish2025complete} (SI Section~S2); node ages are not exact divergence times.}}
\label{sfig:tree}
\end{figure}

We focus on the dataset of Darwin's Finches and their relatives from~\cite{tokita2017cranial,al2021geometry,mosleh2023beak}, which comprises $N = 100$ micro-CT skull specimens drawn from 24 avian species spanning Darwin's Finches radiation and its closest relatives (Fig.~\ref{sfig:tree}). The core Darwin's Finches clade, including \textit{Geospiza}, \textit{Camarhynchus}, \textit{Certhidea}, \textit{Platyspiza}, and \textit{Pinaroloxias}, contributes 53 specimens across 14 species. The remaining 47 specimens span 10 species of close relatives {in the subfamily Coerebinae of the} \textit{Thraupidae} (tanagers){~\cite{burns2014phylogenetics} and provide} phylogenetically distant reference points that anchor the latent space geometry learned by $\mathcal{L}_\text{phylo}$. We apply the voxelization and smoothing procedure as described in~\cite{mosleh2023beak} to remesh the original 3D scans (see Fig.~\ref{FIG_introduction}(a)){, and each skull is centred on its vertex mean and scaled so that its largest absolute coordinate is 1, which removes absolute size}. The dataset is severely class-imbalanced: specimen counts per species range from 2 (\textit{G.~difficilis}, {\textit{G.~scandens},} \textit{G.~septentrionalis}, \textit{P.~inornata}, \textit{C.~fusca}) to 7 (\textit{G.~conirostris}), with a mean of 4.2 specimens per species{, reflecting} the genuine scarcity of museum specimens for rare or island-endemic species.

\begin{figure}[h!]
\centering
\includegraphics[width=\linewidth]{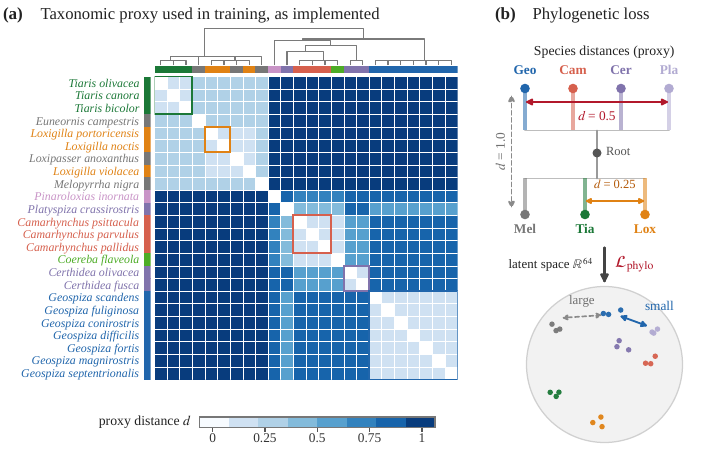}
\caption{\textbf{The phylogenetic structure that drives the phylogenetic consistency loss in our formulation.} (a)~Normalized pairwise phylogenetic distance matrix $d_\text{phylo}$ {as implemented for the taxonomic proxy (SI Table~S2), ordered by average-linkage clustering, with genera outlined}. (b)~The phylogenetic consistency loss $\mathcal{L}_\text{phylo}$ maps {supplied species} distances into the regularized latent space $\mathbb{R}^{64}${; SI Fig.~S1 compares the proxy with the dated tree}.}
\label{sfig:phylo}
\end{figure}

{Species distances are patristic distances on the dated avian tree of McTavish et al.~\cite{mctavish2025complete}, pruned to the 24 species (0.19--16.6\,Myr). A separate set of experiments uses a genus-level taxonomic proxy based on}~\cite{burns2014phylogenetics, burns2002phylogenetic} { We found that genus-parsing errors and an order-dependent lookup had changed 63 of its 276 species pairs as implemented (for example, \textit{Coereba} was read as \textit{Camarhynchus}), leaving a correlation of $r=0.32$ with the tree (SI Fig.~S1 and Table~S2). The main sampler comparison therefore uses a tree-trained representation, and the withheld-species test compares supervision choices and tree-based prediction. The component studies and the morphometric audit use the representation trained with the implemented proxy (Fig.~\ref{sfig:phylo}); every comparison is made within one representation.}

\subsection{Baselines} \label{sec:baselines}

We compare the proposed Residual CFM against {simpler samplers that share its decoder, fitted codes, species centroids and generation slots, and so differ only in how residuals are drawn. The \emph{pooled Gaussian residual} sampler fits the full empirical residual covariance, and the \emph{empirical residual} sampler adds a training residual, drawn uniformly from all 100, to the target centroid; both pool residuals across species, as Residual CFM does.} The third is {the} non-neural Gaussian Mixture Model (GMM) fits a per-species Gaussian with {its covariance shrunk toward the global code covariance (SI Section~S4)}. This serves as a strong statistical baseline that {requires no neural training}. Standard CFM {on full latent codes appears with the component variants (Section~\ref{sec:ablation_results}). DDPM, which produced no valid meshes, could not be rerun because its inputs were not recovered (SI Section~S6).}

\subsection{Evaluation Metrics} \label{sec:metrics}

{Mesh-distance metrics} are computed on 10K-point clouds sampled from generated and training meshes {in the decoder frame, averaged over two surface-sampling seeds, with Chamfer values (Eq.~\eqref{eq:cd}) reported $\times10^{-3}$}. We adopt the 3D generative model metrics as follows:
\begin{itemize}
\item \textbf{Fidelity} (NN-CD, lower is better) measures the mean Chamfer Distance from each generated mesh to its nearest training mesh{, read against the copying level and the spacing of real skulls}.
\item \textbf{Coverage} (COV, higher is better) reports the fraction of training shapes {chosen as nearest neighbor by at least one generated shape}.
\item \textbf{Minimum Matching Distance} (MMD-CD, lower is better) {averages each real mesh's distance to its closest generated mesh, first within each species and then equally across species. In the 24-species mixture, each species has weight $1/24$, shared equally by its specimens (SI Section~S3).}
\item {\textbf{1-NNA} is the leave-one-out accuracy of a nearest-neighbor classifier on the union of generated and real meshes (0.5 at chance).}
\item \textbf{Intra-species diversity}, defined as the pairwise CD among generated samples {of one species} expressed as a percentage of the real intra-species CD.
\item \textbf{Latent Phylogenetic Consistency} (LPC) is the Pearson correlation between {pairwise} latent and {supervising} distances{; agreement with an unused matrix uses a one-sided Mantel test~\cite{mantel1967detection} on species centroids}.
\item \textbf{Morphometric Fr\'{e}chet Distance} ({MorphFID-8}, lower is better) computes a Fr\'{e}chet distance on {eight} morphometric shape descriptors (surface area, volume, bounding box dimensions, elongation, flatness, compactness).
\end{itemize}
{The main comparison uses one representation trained on all 100 skulls, with flow training and baseline sampling repeated for seeds 42--44 (means and sample SD reported). Each sampler and seed fills the same 140 slots: a species mixture of 100 skulls across the 24 species and ten for each of four focal species.}

\subsection{Main Results} \label{sec:results}

\begin{figure}[h!]
\centering
\includegraphics[width=1.0\linewidth]{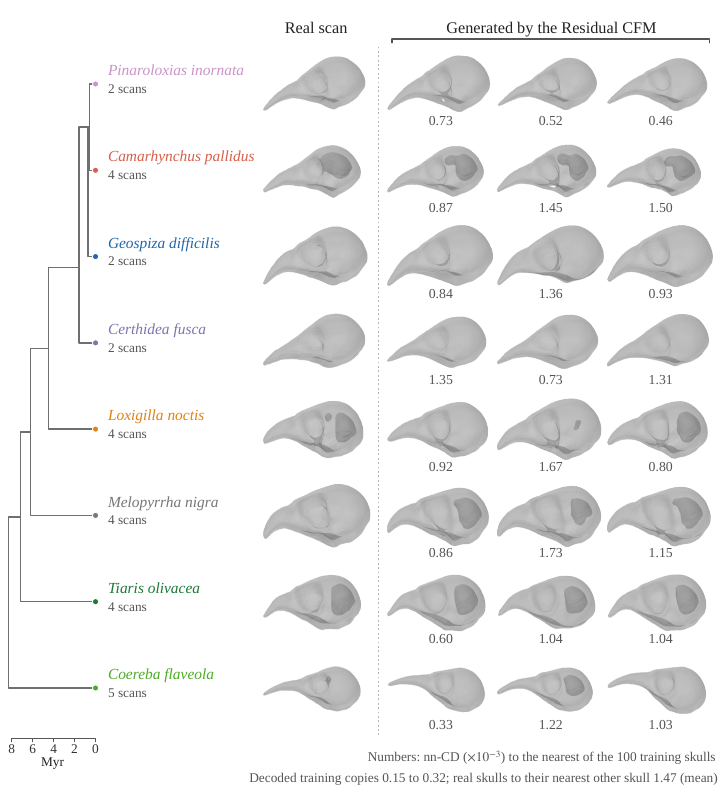}
\caption{{\textbf{Skulls generated across the radiation.} One species per clade (fewest scans), arranged on the dated tree: first real scan and first three seed-42 mixture outputs, in slot order without selection. Labels give nn-CD ($\times10^{-3}$) to the 100 training skulls; decoded-copy range, 0.15--0.32; real leave-one-out mean, 1.47. Lateral views at $256^3$, common normalized scale.}}
\label{fig:generated_grid}
\end{figure}

\begin{figure}[h!]
    \centering
    \includegraphics[width=0.9\linewidth]{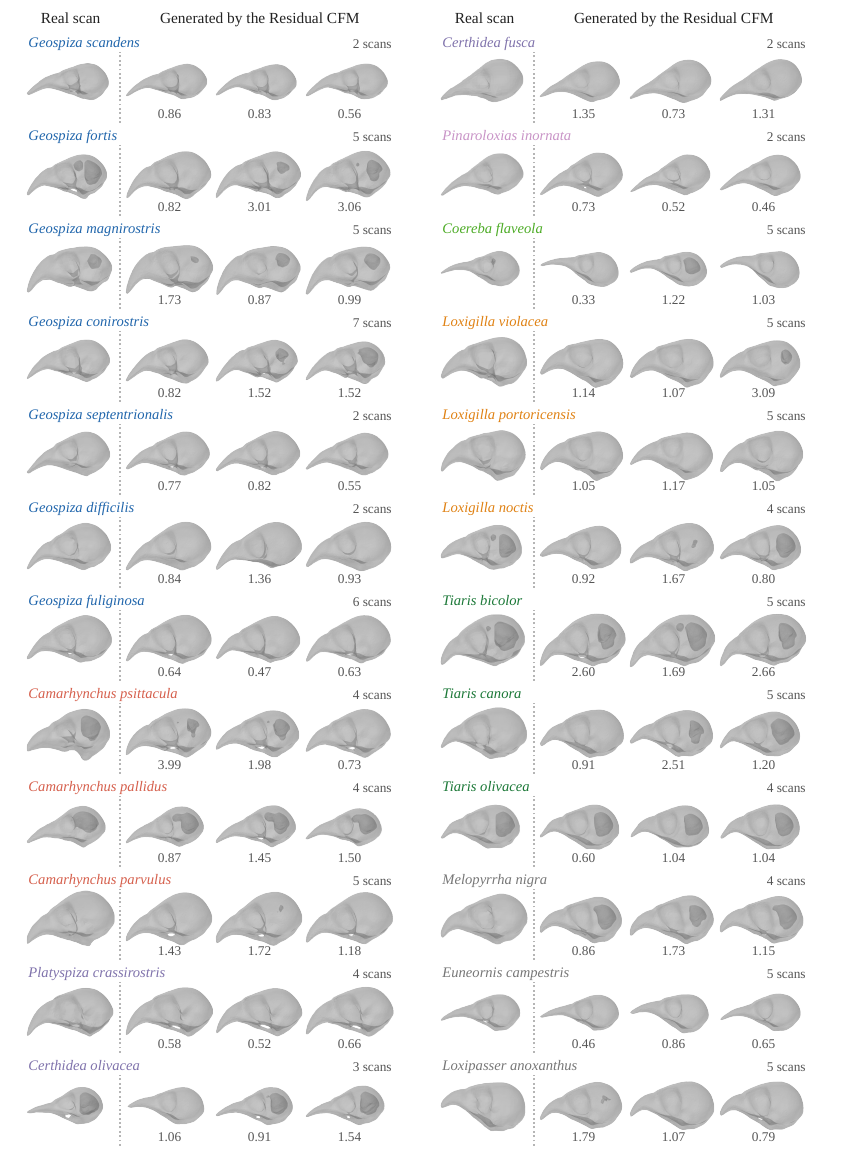}
    \caption{{\textbf{Skulls generated by the flow for all 24 species} (dated-tree comparison, seed 42). Each block shows the first scan of the species and the first three skulls of that species in the species mixture, in slot order and without selection, with nn-CD ($\times10^{-3}$) to the nearest of the 100 training skulls. Species are grouped by clade as in main-text Fig.~3 and follow the dated tree within each clade. Main-text Fig.~5 shows one species per clade.}}
    \label{sfig:tree_gallery}
\end{figure}

Figure~\ref{fig:generated_grid} shows {a real skull and three generated skulls for one species per clade, including species with only two scans. These views illustrate variation in skull shape.}
{Fig.~\ref{sfig:tree_gallery} shows, for every species, its first scan and the first three species-mixture skulls of the flow in seed 42.}

{All three pooled residual samplers improve on per-species sampling (GMM) in both mixture distances (Table~\ref{tab:main}). Residual CFM has the lowest mean species-mixture MMD-CD over three seeds: 1.175, approximately 2\% below empirical residual sampling and 16.5\% below the GMM. It beats both Gaussian samplers in every seed and empirical residual sampling in two of three, with the smallest between-seed SD (0.005 versus 0.045--0.120). These are descriptive comparisons on one fixed representation; differences among samplers were not tested for significance.}

{Residual CFM also has the lowest nn-CD in every seed. Its mean (1.203) is below real-skull spacing (1.467) and above decoded-copy proximity (0.234). Its 1-NNA lies within the 95\% random-relabelling interval (about 0.41--0.58) in every seed, as do those of the other pooled samplers. A separate proxy-based comparison gives the same mean MMD-CD ordering, but paired tests detect an advantage only over the GMM, and Residual CFM's 1-NNA is above chance there (SI Section~S5).}

\begin{table}[t]
\centering
\footnotesize
\setlength{\tabcolsep}{3pt}
\resizebox{\linewidth}{!}{%
{\color{black}
\begin{tabular}{lccccc}
\toprule
\textbf{Sampler} & \textbf{nn-CD}$\downarrow$ & \textbf{MMD-CD}$\downarrow$ & \textbf{COV}$\uparrow$ & \textbf{1-NNA}$\to0.5$ & \textbf{MorphFID-8}$\downarrow$\\
\midrule
Residual CFM (ours) & 1.203\,$\pm$\,0.039 & 1.175\,$\pm$\,0.005 & 0.513\,$\pm$\,0.015 & 0.550\,$\pm$\,0.018 & 1.674\,$\pm$\,0.244\\
Pooled Gaussian residual & 1.328\,$\pm$\,0.030 & 1.246\,$\pm$\,0.045 & 0.543\,$\pm$\,0.051 & 0.533\,$\pm$\,0.028 & 1.102\,$\pm$\,0.314\\
Empirical residual & 1.381\,$\pm$\,0.040 & 1.198\,$\pm$\,0.063 & 0.532\,$\pm$\,0.033 & 0.519\,$\pm$\,0.038 & 0.849\,$\pm$\,0.070\\
GMM (per-species Gaussian) & 1.582\,$\pm$\,0.119 & 1.408\,$\pm$\,0.120 & 0.593\,$\pm$\,0.010 & 0.473\,$\pm$\,0.069 & 0.973\,$\pm$\,0.211\\
\midrule
Real skulls (leave-one-out) & 1.467 & n/a & n/a & n/a & n/a\\
\bottomrule
\end{tabular}
}}
\caption{\textbf{Main quantitative results.} {Sampler comparison on the tree-trained representation (species mixture, 100 generated and 100 real skulls per seed; means $\pm$ SD over seeds 42--44; Chamfer values $\times10^{-3}$). The last row compares each real skull with the other 99 (n/a, not applicable). MorphFID-8 uses convex-hull volume for open outputs. Per-seed values and focal-species metrics are in SI Section~S4.}}
\label{tab:main}
\end{table}

\begin{figure}[h!]
\centering
\includegraphics[width=1\linewidth]{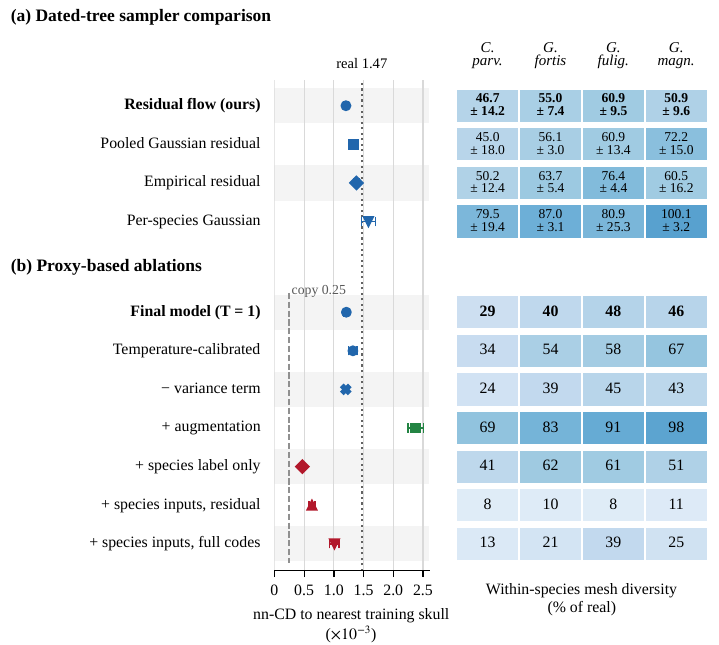}
\caption{{\textbf{Fidelity and within-species mesh diversity.} (a)~Tree-trained main comparison (Table~\ref{tab:main}; ``Residual flow'' is Residual CFM and the per-species Gaussian is the GMM), means and SD over three seeds. (b)~Proxy-based Residual CFM ablations (SI Table~S15). Columns are \textit{C.~parvulus}, \textit{G.~fortis}, \textit{G.~fuliginosa} and \textit{G.~magnirostris}; dotted line, real-skull spacing; dashed line, copying reference for panel~(b).}}
\label{fig:fidelity}
\end{figure}

{The GMM spreads its samples more widely: it reaches the highest coverage (0.593 against 0.513), retains 79.5--100.1\% of real within-species mesh diversity in the four focal species against 46.7--60.9\% for Residual CFM (Fig.~\ref{fig:fidelity}a), and has lower focal-species MMD-CD in three of the four species (SI Table~S6). Residual CFM thus trades within-species spread for lower mixture distances, a balance that the sampling temperature can shift (Section~\ref{sec:ablation_results}).}

\subsection{Generativity Verification} \label{sec:verification}

{We use three complementary diagnostics to examine training-set proximity and the residual signal (Table~\ref{tab:verification}). Together they characterize the outputs but do not establish that memorization is absent.}

\paragraph{Test 1: Memorization Check}
For each of the {300 species-mixture meshes generated by Residual CFM, and those of the three baselines}, we compute its Chamfer Distance to all 100 training meshes. A generated mesh is flagged as ``memorized'' if its distance to the nearest training mesh falls below 10\% of the minimum same-species training distance {($0.026\times10^{-3}$)}. Of {1,200} generated meshes, zero fall below this threshold{, but neither do the 16 exact training-code copies from empirical resampling. Against the range of decoded training codes (0.153--0.320), only one Residual CFM output (0.272) comes as close.}

\paragraph{Test 2: Residual Meaningfulness}
This test asks whether the {training} residuals {are} negligibly small{, which would make the representation} effectively a centroid-lookup table. {On the proxy-based representation, the} mean residual magnitude $\mathbb{E}[\|\br_i\|_2] = 0.380$ {is 42.0\% of} the mean inter-species centroid distance $\mathbb{E}[\|\bmu_a-\bmu_b\|_2] = 0.904${, so the residuals give the flow substantial within-species variation to learn; this describes the representation, not the diversity of generated meshes}.

\paragraph{Test 3: Mesh-level Novelty Ratio}
We define the novelty ratio {$\rho = \mathrm{CD}_{\mathrm{NN}} / \mathrm{CD}_{\mathrm{sp}}$}, where the numerator is the mean Chamfer Distance from each generated mesh to its nearest training mesh and the denominator is the mean pairwise CD among real same-species specimens. {For Residual CFM, $\rho = 21.4\%$, between decoded training codes ($4.2\%$) and real skulls ($26.1\%$, from each skull's nearest other skull). This ratio measures proximity to the training set rather than diversity and does not, by itself, establish generativity.}

\begin{table}[t]
\centering
\footnotesize
\begin{tabular}{llc}
\toprule
\textbf{Test} & \textbf{Metric} & \textbf{Value} \\
\midrule
1.\ Memorization & {Flagged; within copy range} & {0\,/\,300; 1\,/\,300} \\
2.\ Residual signal & $\|\br\|\,/\,\|\bmu_a-\bmu_b\|$ & 42.0\% \\
3.\ Mesh novelty & {$\mathrm{CD}_{\mathrm{NN}}/\mathrm{CD}_{\mathrm{sp}}$} & {21.4\% (copy 4.2\%, real 26.1\%)} \\
\bottomrule
\end{tabular}
\caption{\textbf{Generativity verification.} {Complementary diagnostics that do not by themselves establish the absence of memorization. Tests 1 and 3 use the 300 species-mixture outputs of Residual CFM in Table~\ref{tab:main}; Test 2 uses the proxy-based representation.}}
\label{tab:verification}
\end{table}

\subsection{Ablation Study} \label{sec:ablation_results}

{SI Table~S15 ablates the final Residual CFM over three seeds on the proxy-based representation (Fig.~\ref{fig:fidelity}b); SI Table~S14 removes components from the initial conditioned design. Adding species, size and phylogeny inputs lowers mixture distances but reduces within-species diversity to 8--11\% of real. Species labels alone retain 41--62\% diversity but yield copy-like proximity (nn-CD 0.47). These results motivate the neutral inputs of the final flow.}

{The initial design expanded 100 codes to 1,107 through perturbation, interpolation and centroid extrapolation (SI Section~S6). Training the final flow on these augmented codes raises diversity to 69--98\%, but also raises nn-CD from 1.21 to 2.38 and MMD-CD from 1.24 to 1.77 (both $p=0.0005$), so the final model uses real codes only. Standard CFM on full codes improves on the initial design (nn-CD 1.57 versus 2.07, $p=0.002$) and retains 42--66\% of real within-species diversity.}

{Temperature calibration ($T=1.79$--$1.85$) raises COV by 0.06 ($p=0.006$) and diversity to 34--67\%, at an nn-CD cost of 0.11 ($p=0.02$). Removing the variance penalty has no detectable effect ($p\ge0.53$); the ablation therefore provides no evidence that it improves diversity.}

\subsection{{Phylogenetic Supervision}} \label{sec:phylo_results}

{We retrained the representation without the distance loss, with either matrix and with three weakly correlated shuffled proxies (SI Section~S7). Every supervised representation fits its target (LPC 0.99, against 0.23 for the proxy without the loss) without worsening reconstruction, so LPC measures fit to whichever matrix supervises training. For species seen in training the matrix matters little to generation, apart from a focal-species MMD-CD gain with the proxy (1.81 to 1.47, $p=0.008$; SI Table~S17; $p=0.036$  in the conditional-CFM analysis of SI Section~S10) that the shuffled controls do not tie to correct relationships. The value of the tree appears when species are missing from training.}

\subsection{Leave-One-Species-Out Generalization} \label{sec:loso}

{A leave-one-species-out (LOSO) analysis estimated each centroid from the three nearest species by inverse-distance weighting, across the 18 species with at least four specimens. Prediction error averaged 0.37 against a mean within-species spread of 0.62, and was below the spread for every species (SI Section~S10). Because representation learning had used all specimens, this describes trained latent geometry; the retrained holdout below tests prediction without withheld scans.}

\subsection{{Withheld Species}} \label{sec:heldout}

{We retrained the representation on 79 skulls from 20 species, withholding all 21 skulls of the four focal species, and retrained Residual CFM on training residuals (one split; SI Section~S8). Without using withheld scans, we added identical sampled residuals to centroids predicted from three neighbouring training species or to the global mean of training centroids.}

\begin{figure}[h!]
\centering
\includegraphics[width=1\linewidth]{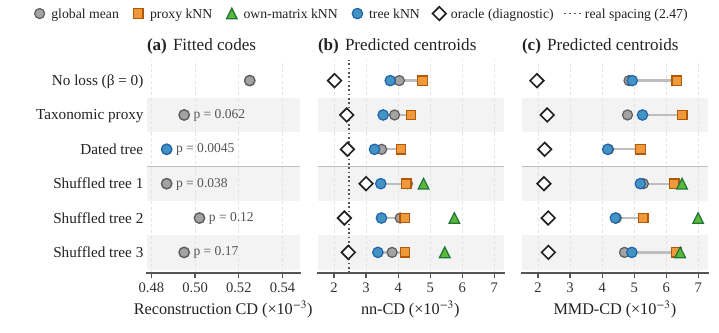}
\caption{{\textbf{Withheld species} (79 training skulls from 20 species; all 21 skulls of four focal species withheld; one split). (a)~Reconstruction after fitting codes to the withheld skulls, with $p$ from the paired sign-flip test against no distance loss. (b,~c)~Generation from non-oracle centroids (global mean, or three neighbours in the proxy, the training matrix or the dated tree) and from the oracle diagnostic; grey segments pair proxy and tree predictions. Values are in SI Tables~S18 and~S20.}}
\label{fig:heldout}
\end{figure}

{Dated-tree neighbours lower nn-CD relative to the global mean in all six representations and MMD-CD in three (Fig.~\ref{fig:heldout}b,c). With the tree-trained representation, nn-CD reaches 3.26, about 1.3 times real-conspecific spacing (2.47). Much of the gain comes from neighbour choice: shuffled-trained representations reach 3.37--3.48, and tree neighbours beat proxy neighbours in all six representations, mainly because the proxy misplaces \textit{C.~parvulus} (SI Table~S19).}

{Separately, we fitted codes to the withheld scans with the decoder frozen and used their centroids as an oracle diagnostic. Its nn-CD of 2.42 is close to real-conspecific spacing, showing how fitted codes can extend the model when scans become available. Tree supervision reduces reconstruction CD from 0.525 without the loss to 0.487 ($p=0.0045$; Fig.~\ref{fig:heldout}a); shuffled matrices also help (0.487--0.502). The oracle uses withheld geometry and is distinct from tree-based prediction. A single split does not establish transfer across other species selections.}

\subsection{Morphometric Fr\'{e}chet Distance} \label{sec:mfid}

We introduce \textbf{MorphFID{-8}}, a Fr\'{e}chet distance computed in the space of biologically interpretable morphometric descriptors rather than learned image features{: the eight descriptors of Section~\ref{sec:metrics}, $z$-scored with species-balanced training statistics, with Eq.~\eqref{eq:frechet} applied to their Gaussian fits. A 12-descriptor variant additionally includes the vertex count and three centroid coordinates (SI Table~S12).  The vertex count depends on grid resolution and dominates that 12-dimensional score (SI Table~S12), so we omit these four. Residual CFM has the largest MorphFID-8 (1.674, against 0.849 for empirical residuals), consistent with inflated convex-hull volumes for its open outputs, nearly all of which contact one face of the extraction box. On the six descriptors that do not use the hull, it agrees best with the real skulls in every seed, and its closed-output MorphFID-8 is 0.559. This subset comparison is exploratory because selection and sample sizes differ across samplers (SI Table~S8).}

\subsection{{Morphometric Audit}} \label{sec:anatomy}

\begin{figure}[h!]
\centering
\includegraphics[width=0.90\linewidth]{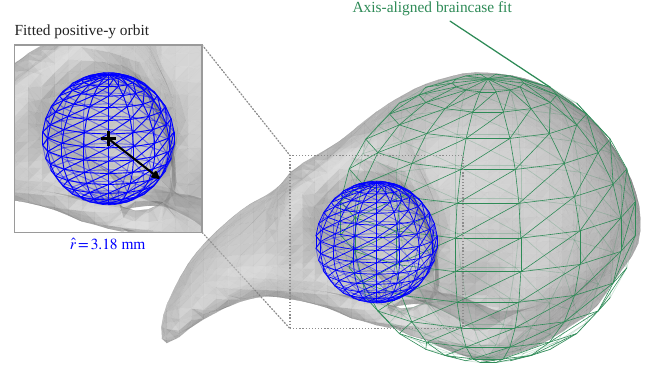}
\caption{{\textbf{Orbit and braincase fits on a generated skull of \textit{C. parvulus}.} A sphere (blue) and an axis-aligned ellipsoid (green) fitted with the method of~\cite{lau2025robust} to a generated \textit{C.~parvulus} skull (proxy-based model, $T=1$, seed~42, slot~000). The inset shows the positive-$y$ orbit, with fitted radius 3.18\,mm.}}
\label{sfig:anatomy_fits}
\end{figure}

\begin{figure}[h!]
\centering
\includegraphics[width=1\linewidth]{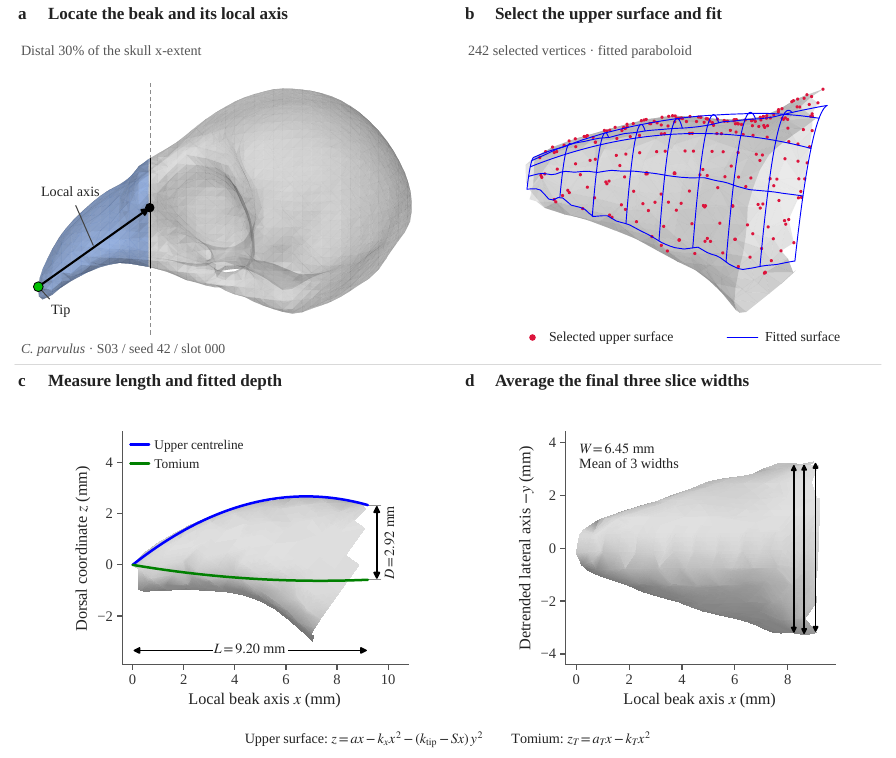}
\caption{{\textbf{Automatic beak measurement} on the skull of Fig.~\ref{sfig:anatomy_fits}. (a)~The distal 30\% of skull length (a geometric convention) is selected and a local beak axis defined. (b)~The upper surface is selected and a paraboloid fitted~\cite{al2021geometry,mosleh2023beak,lau2025robust}. (c)~Tip-to-base length $L$ and fitted depth $D$ are measured at the base. (d)~Width $W$ averages the final three valid axial slices.}}
\label{sfig:beak_procedure}
\end{figure}

{Mesh distances do not show whether generated skulls keep the proportions that distinguish species. In a separate audit, we measured 12 size-free skull, orbit, braincase and beak proportions on 420 proxy-based Residual CFM outputs ($T=1$, three seeds) and on decoded real skulls as reference, using the sphere and ellipsoid fits of~\cite{lau2025robust} for orbits and braincases (Fig.~\ref{sfig:anatomy_fits}) and the paraboloid model of~\cite{al2021geometry,mosleh2023beak} for beaks (Fig.~\ref{sfig:beak_procedure}; SI Section~S9).}

\begin{figure}[h!]
\centering
\includegraphics[width=1\linewidth]{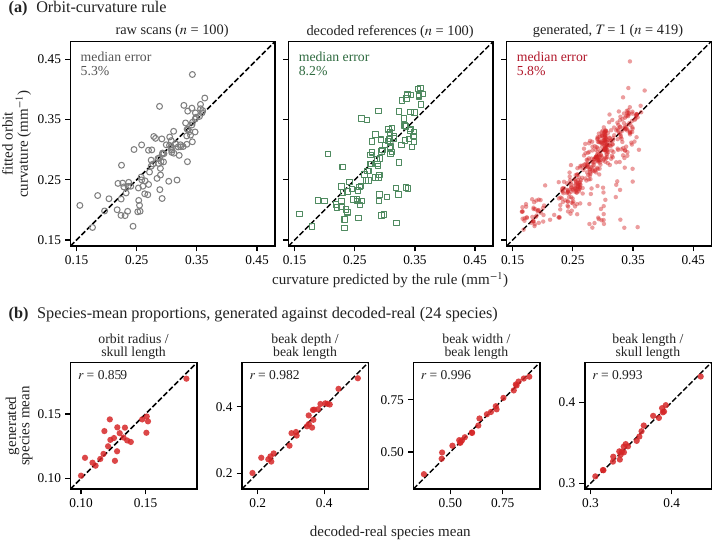}
\caption{{\textbf{Automated morphometry of proxy-based Residual CFM outputs.} (a)~Fitted orbit curvature against the rule of~\cite{lau2025robust} for 100 raw scans, 100 decoded real skulls and 419 complete generated skulls, with median relative errors. (b)~Generated against decoded-real species means for four proportions across 24 species, with correlations and the identity line dashed.}}
\label{fig:morphometry}
\end{figure}

{Generated and decoded-real species means correlate at $r=0.982$ for beak depth/beak length, $r=0.996$ for beak width/beak length and $r=0.993$ for beak length/skull length (Fig.~\ref{fig:morphometry}b), with a median of 0.98 across all 12 measures. The skulls also follow the orbit-curvature rule of~\cite{lau2025robust}, which played no part in training (median error 5.8\%, against 5.3\% for raw scans and 8.2\% for decoded real skulls; Fig.~\ref{fig:morphometry}a), and 98.5\% of the 5,028 values from the 419 complete skulls lie within two pooled reference SD of the species mean (reference 97.2\%). This agreement matters because finch species are distinguished by such proportions, and species identity enters generation only through the centroid. Within-species spread is narrower than in real skulls (55\% of the decoded-real spread, 73\% with temperature calibration), matching the mesh diversity. These measurements describe proxy-based outputs, not the tree-trained samples of Table~\ref{tab:main}, and are not an expert anatomical assessment.}

\section{Discussion} \label{sec:discussion}

\subsection{Why Residual Decomposition Works}\label{sec:why_pooling}
{Species centroids separate between-species location from residual variation, so all 100 specimens can inform a shared within-species distribution. This random-effects structure shares evidence across species rather than creating more independent observations. The matched comparison supports pooling: every pooled sampler improves on per-species sampling in both mixture distances (Table~\ref{tab:main}). A common residual distribution remains a working assumption; whether it suits more distant taxa or other grouped shape datasets requires testing.}

\subsection{Role of Phylogenetic Consistency Loss}\label{sec:role_phylo}
While the residual decomposition addresses the sample-size bottleneck, the phylogenetic consistency loss addresses a complementary structural bottleneck. Without this regularization, the DeepSDF latent space is organized purely by reconstruction fidelity, and there is no guarantee that the resulting code geometry reflects evolutionary relationships. {Our controls show that each tested distance matrix lowers reconstruction error for unseen skulls, whereas the phylogenetic tree supports} prediction {for species absent from training} via phylogenetic neighbor weighting{. A species-level tree simplifies relationships complicated by hybridization among ground finches~\cite{lamichhaney2015evolution}, and these results concern extant species rather than ancestral reconstruction}.

\subsection{Neural vs.\ Statistical Generation}\label{sec:neural_vs_stat}
{The GMM retains more within-species diversity, while Residual CFM has a mean mixture MMD-CD 16.5\% lower and more stable across the three seeds. Much of the gain over the GMM is shared by simpler pooled samplers; the additional mean improvement over empirical residual sampling is approximately 2\%. The neural sampler's flexibility could help on larger or multimodal datasets, for example with sexual dimorphism~\cite{frayer1985sexual}, but this is a hypothesis for future testing.}

{The present tradeoff is narrower within-species spread. In the proxy-based diagnostics, decoding training codes retains 98--100\% of real within-species diversity, whereas sampled residual variance is 0.48--0.57 of training variance (SI Section~S5). This implicates the generator without excluding decoder effects at generated codes.}

\subsection{Limitations}\label{sec:limitations}
Several limitations of the current work merit discussion. The {dated tree} relies on branch lengths from molecular phylogenies, which may not perfectly reflect morphological divergence rates, particularly for lineages that have undergone rapid adaptive radiation or convergent evolution. {The proxy-based studies carry the documented errors of the taxonomic proxy, and the main comparison uses one representation, so its seeds measure variation in flow training and sampling only.} The coverage score of {0.513 and the within-species diversity of 47--61\%} {indicate} that the model does not yet span the full training distribution. {Across all four samplers in the main comparison, 624 of 1,680 outputs are non-watertight, and 623 contact the extraction boundary (SI Table~S7); re-extraction on a larger domain is needed to test whether these surfaces close.} Finally, the biological plausibility of generated specimens requires expert morphological evaluation by comparative anatomists, which we leave to future collaborative work.

\subsection{Broader Impact}
{PhyloSDF provides a way to explore synthetic morphological variation when scans are scarce. Whether synthetic specimens improve downstream morphometric inference remains to be tested; they do not replace independent biological observations. Outputs should be labelled as synthetic, with expert assessment preceding biological applications.}

\section{Conclusion} \label{sec:conclusion}

{PhyloSDF generates complete 3D skulls from 100 specimens across 24 species with only 2--7 scans per species, combining phylogenetic regularization, an implicit shape representation and a learned residual flow. Species-centroid decomposition lets specimens from different species inform within-species variation. On a fixed tree-trained representation, Residual CFM achieves the lowest mean species-mixture MMD-CD over three seeds, with a modest additional gain over empirical residual sampling. A separate morphometric audit shows preservation of measured species-level proportions. Dated-tree neighbours support prediction without withheld scans, and fitted codes extend the model when scans become available.}

{The immediate priorities are to recover more within-species variation, resolve open surfaces and obtain expert anatomical assessment. Larger datasets and multiple species splits can test whether the residual framework remains useful as sampling grows and variation becomes more complex. The present results establish a practical approach to structured 3D generation under severe specimen scarcity and a basis for those tests.}

\noindent {\textbf{Data Availability}} \ \ { The code and data are available at \url{https://github.com/kaikwanlau/PhyloSDF}.}

\bibliographystyle{ieeetr}
\bibliography{references.bib}

@article{cunningham2014virtual,
  title = {A virtual world of paleontology},
  volume = {29},
  ISSN = {0169-5347},
  number = {6},
  journal = {Trends Ecol. Evol.},
  publisher = {Elsevier BV},
  author = {Cunningham,  John A. and Rahman,  Imran A. and Lautenschlager,  Stephan and Rayfield,  Emily J. and Donoghue,  Philip C.J.},
  year = {2014},
  pages = {347--357}
}

@article{davies2017open,
  title = {Open data and digital morphology},
  volume = {284},
  ISSN = {1471-2954},
  number = {1852},
  journal = {Proc. R. Soc. B},
  publisher = {The Royal Society},
  author = {Davies,  Thomas G. and Rahman,  Imran A. and Lautenschlager,  Stephan and Cunningham,  John A. and Asher,  Robert J. and Barrett,  Paul M. and Bates,  Karl T. and Bengtson,  Stefan and Benson,  Roger B. J. and Boyer,  Doug M. and others},
  year = {2017},
  pages = {20170194}
}

@article{klingenberg2010evolution,
  title = {Evolution and development of shape: integrating quantitative approaches},
  volume = {11},
  ISSN = {1471-0064},
  number = {9},
  journal = {Nat. Rev. Genet.},
  publisher = {Springer Science and Business Media LLC},
  author = {Klingenberg,  Christian Peter},
  year = {2010},
  pages = {623--635}
}

@article{mosleh2023beak,
  title = {Beak morphometry and morphogenesis across avian radiations},
  volume = {290},
  ISSN = {1471-2954},
  number = {2007},
  journal = {Proc. R. Soc. B},
  publisher = {The Royal Society},
  author = {Mosleh,  Salem and Choi,  Gary P. T. and Musser,  Grace M. and James,  Helen F. and Abzhanov,  Arhat and Mahadevan,  L.},
  year = {2023},
  pages = {20230420}
}

@article{keklikoglou2019micro,
  title = {Micro-computed tomography for natural history specimens: a handbook of best practice protocols},
  ISSN = {2118-9773},
  number = {522},
  pages = {1--55},
  journal = {Eur. J. Taxon.},
  publisher = {Museum National D'Histoire Naturelle},
  author = {Keklikoglou,  Kleoniki and Faulwetter,  Sarah and Chatzinikolaou,  Eva and Wils,  Patricia and Brecko,  Jonathan and Kva{\v{c}}ek,  Ji{\v{r}}{\'i} and Metscher,  Brian and Arvanitidis,  Christos},
  year = {2019}
}

@article{bardua2019practical,
  title={A practical guide to sliding and surface semilandmarks in morphometric analyses},
  volume = {1},
  ISSN = {2517-4843},
  number = {1},
  pages = {obz016},
  journal = {Integr. Org. Biol.},
  publisher = {Oxford University Press (OUP)},
  author = {Bardua,  C and Felice,  R N and Watanabe,  A and Fabre, Anne-Claire and Goswami,  A},
  year = {2019}
}

@article{mulqueeney2024many,
  title={How many specimens make a sufficient training set for automated three-dimensional feature extraction?},
  volume = {11},
  ISSN = {2054-5703},
  number = {6},
  journal = {R. Soc. Open Sci.},
  publisher = {The Royal Society},
  author = {Mulqueeney,  James M. and Searle-Barnes,  Alex and Brombacher,  Anieke and Sweeney,  Marisa and Goswami,  Anjali and Ezard,  Thomas H. G.},
  year = {2024},
  pages = {rsos.240113}
}

@article{goodfellow2020generative,
  title={Generative adversarial networks},
  volume = {63},
  ISSN = {1557-7317},
  number = {11},
  journal = {Commun. ACM},
  publisher = {Association for Computing Machinery (ACM)},
  author = {Goodfellow,  Ian and Pouget-Abadie,  Jean and Mirza,  Mehdi and Xu,  Bing and Warde-Farley,  David and Ozair,  Sherjil and Courville,  Aaron and Bengio,  Yoshua},
  year = {2020},
  pages = {139--144}
}

@article{ho2020denoising,
  title={Denoising diffusion probabilistic models},
  author={Ho, Jonathan and Jain, Ajay and Abbeel, Pieter},
  journal={NeurIPS},
  volume={33},
  pages={6840--6851},
  year={2020}
}

@inproceedings{lipman2022flow,
    title = {Flow Matching for Generative Modeling},
    author = {Lipman, Yaron and Chen, Ricky T. Q. and Ben-Hamu, Heli and Nickel, Maximilian and Le, Matt},
    year = {2023},
    booktitle = {ICLR} 
}

@article{van2021memorization,
  title={On memorization in probabilistic deep generative models},
  author={van den Burg, Gerrit and Williams, Chris},
  journal={NeurIPS},
  volume={34},
  pages={27916--27928},
  year={2021}
}

@article{zhang2021understanding,
  title={Understanding deep learning (still) requires rethinking generalization},
  volume = {64},
  ISSN = {1557-7317},
  number = {3},
  journal = {Commun. ACM},
  publisher = {Association for Computing Machinery (ACM)},
  author = {Zhang,  Chiyuan and Bengio,  Samy and Hardt,  Moritz and Recht,  Benjamin and Vinyals,  Oriol},
  year = {2021},
  pages = {107--115}
}

@article{felsenstein1985phylogenies,
  title={Phylogenies and the comparative method},
  volume = {125},
  ISSN = {1537-5323},
  number = {1},
  journal = {Am. Nat.},
  publisher = {University of Chicago Press},
  author = {Felsenstein,  Joseph},
  year = {1985},
  pages = {1--15}
}

@inproceedings{chong2020effectively,
  title={Effectively Unbiased {FID} and Inception Score and Where to Find Them}, 
  author={Chong, Min Jin and Forsyth, David},
  booktitle={CVPR}, 
  year={2020},
  volume={},
  number={},
  pages={6069--6078},
  }

@inproceedings{park2019deepsdf,
  title={{DeepSDF}: Learning Continuous Signed Distance Functions for Shape Representation}, 
  author={Park, Jeong Joon and Florence, Peter and Straub, Julian and Newcombe, Richard and Lovegrove, Steven},
  booktitle={CVPR}, 
  year={2019},
  volume={},
  number={},
  pages={165--174},
  }

@inproceedings{gropp2020implicit,
  title={Implicit geometric regularization for learning shapes},
  author={Gropp, Amos and Yariv, Lior and Haim, Niv and Atzmon, Matan and Lipman, Yaron},
  booktitle={ICML},
  pages={3789--3799},
  year={2020}
}

@inproceedings{lorensen1998marching,
  title={Marching cubes: A high resolution {3D} surface construction algorithm},
  author={Lorensen, William E and Cline, Harvey E},
  booktitle={Seminal graphics: Pioneering efforts that shaped the field},
  pages={347--353},
  year={1998}
}

@inproceedings{albergo2022building,
  title={Building normalizing flows with stochastic interpolants},
  author={Albergo, Michael S and Vanden-Eijnden, Eric},
  booktitle={ICLR},
  year={2023}
}

@article{burns2014phylogenetics,
  title={Phylogenetics and diversification of tanagers ({P}asseriformes: {T}hraupidae), the largest radiation of {N}eotropical songbirds},
  author={Burns, Kevin J and Shultz, Allison J and Title, Pascal O and Mason, Nicholas A and Barker, F Keith and Klicka, John and Lanyon, Scott M and Lovette, Irby J},
  journal={Mol. Phylogenet. Evol.},
  volume={75},
  pages={41--77},
  year={2014},
  publisher={Elsevier}
}

@article{burns2002phylogenetic,
  title={Phylogenetic relationships and morphological diversity in {D}arwin's finches and their relatives},
  author={Burns, Kevin J and Hackett, Shannon J and Klein, Nedra K},
  journal={Evolution},
  volume={56},
  number={6},
  pages={1240--1252},
  year={2002},
  publisher={Blackwell Publishing Ltd Oxford, UK}
}

@article{tong2023improving,
    title = "Improving and generalizing flow-based generative models with minibatch optimal transport",
    author = "Alexander Tong and Kilian Fatras and Nikolay Malkin and Guillaume Huguet and Yanlei Zhang and Jarrid Rector-Brooks and Guy Wolf and Yoshua Bengio",
    year = "2024",
    language = "English",
    pages = "1--34",
    journal = "Trans. Mach. Learn. Res.",
    issn = "2835-8856",
}

@article{dowson1982frechet,
  title={The {F}r{\'e}chet distance between multivariate normal distributions},
  author={Dowson, David C and Landau, B V},
  journal={J. Multivar. Anal.},
  volume={12},
  number={3},
  pages={450--455},
  year={1982},
  publisher={Elsevier}
}

@inproceedings{mescheder2019occupancy,
  title={Occupancy Networks: Learning {3D} Reconstruction in Function Space}, 
  author={Mescheder, Lars and Oechsle, Michael and Niemeyer, Michael and Nowozin, Sebastian and Geiger, Andreas},
  booktitle={CVPR}, 
  year={2019},
  volume={},
  number={},
  pages={4455--4465},
}

@inproceedings{chen2019learning,
    title = { Learning Implicit Fields for Generative Shape Modeling },
    author = { Chen, Zhiqin and Zhang, Hao },
    booktitle = {CVPR},
    year = {2019},
    volume = {},
    ISSN = {},
    pages = {5932--5941},
}

@inproceedings{chibane2020neural,
title = {Neural unsigned distance fields for implicit function learning},
author = {Chibane, Julian and Mir, Aymen and Pons-Moll, Gerard},
year = {2020},
isbn = {9781713829546},
booktitle = {NeurIPS},
articleno = {1816},
numpages = {15},
pages={21638--21652}
}

@inproceedings{niemeyer2020differentiable,
  title={Differentiable Volumetric Rendering: Learning Implicit {3D} Representations Without {3D} Supervision}, 
  author={Niemeyer, Michael and Mescheder, Lars and Oechsle, Michael and Geiger, Andreas},
  booktitle={CVPR}, 
  year={2020},
  volume={},
  number={},
  pages={3504--3515},
}

@inproceedings{achlioptas2018learning,
  title={Learning representations and generative models for {3D} point clouds},
  author={Achlioptas, Panos and Diamanti, Olga and Mitliagkas, Ioannis and Guibas, Leonidas},
  booktitle={ICML},
  pages={40--49},
  year={2018},
}

@inproceedings{yang2019pointflow,
  title={{PointFlow}: {3D} Point Cloud Generation With Continuous Normalizing Flows}, 
  author={Yang, Guandao and Huang, Xun and Hao, Zekun and Liu, Ming-Yu and Belongie, Serge and Hariharan, Bharath},
  booktitle={ICCV}, 
  year={2019},
  volume={},
  number={},
  pages={4540--4549},
  }

@inproceedings{maturana2015voxnet,
  title={{VoxNet}: A {3D} Convolutional Neural Network for real-time object recognition}, 
  author={Maturana, Daniel and Scherer, Sebastian},
  booktitle={IROS}, 
  year={2015},
  volume={},
  number={},
  pages={922--928},
  }

@inproceedings{nash2020polygen,
  title = 	 {{P}oly{G}en: An Autoregressive Generative Model of 3{D} Meshes},
  author =       {Nash, Charlie and Ganin, Yaroslav and Eslami, S. M. Ali and Battaglia, Peter},
  booktitle = 	 {ICML},
  pages = 	 {7220--7229},
  year = 	 {2020},
  volume = 	 {119}
}

@inproceedings{hao2020dualsdf,
  title = {{DualSDF}: Semantic Shape Manipulation Using a Two-Level Representation},
  booktitle = {CVPR},
  publisher = {IEEE},
  author = {Hao,  Zekun and Averbuch-Elor,  Hadar and Snavely,  Noah and Belongie,  Serge},
  year = {2020},
  pages = {7628--7638}
}

@article{vahdat2022lion,
  title={{LION}: Latent point diffusion models for {3D} shape generation},
  author={Zeng, Xiaohui and Vahdat, Arash and Williams, Francis and Gojcic, Zan and Litany, Or and Fidler, Sanja and Kreis, Karsten},
  journal={NeurIPS},
  volume={35},
  pages={10021--10039},
  year={2022}
}

@article{pagel1999inferring,
  title={Inferring the historical patterns of biological evolution},
  author={Pagel, Mark},
  journal={Nature},
  volume={401},
  number={6756},
  pages={877--884},
  year={1999},
  publisher={Nature Publishing Group UK London}
}

@inproceedings{kingma2014adam,
  title={Adam: A method for stochastic optimization},
  author={Kingma, Diederik P and Ba, Jimmy},
  booktitle={ICLR},
  year={2015}
}

@article{tokita2017cranial,
  title={Cranial shape evolution in adaptive radiations of birds: comparative morphometrics of {D}arwin's finches and {H}awaiian honeycreepers},
  author={Tokita, Masayoshi and Yano, Wataru and James, Helen F. and Abzhanov, Arhat},
  journal={Philos. Trans. R. Soc. B: Biol. Sci.},
  volume={372},
  number={1713},
  pages={20150481},
  year={2017},
  publisher={The Royal Society}
}

@article{al2021geometry,
  title={Geometry and dynamics link form, function, and evolution of finch beaks},
  author={Al-Mosleh, Salem and Choi, Gary P. T. and Abzhanov, Arhat and Mahadevan, L.},
  journal={Proc. Natl. Acad. Sci.},
  volume={118},
  number={46},
  pages={e2105957118},
  year={2021},
  publisher={National Academy of Sciences}
}

@article{lamichhaney2015evolution,
  title={Evolution of {D}arwin's finches and their beaks revealed by genome sequencing},
  author={Lamichhaney, Sangeet and Berglund, Jonas and Alm{\'e}n, Markus S{\"a}llman and Maqbool, Khurram and Grabherr, Manfred and Martinez-Barrio, Alvaro and Promerov{\'a}, Marta and Rubin, Carl-Johan and Wang, Chao and Zamani, Neda and Grant, B Rosemary and Grant, Peter R and Webster, Matthew T and Andersson, Leif},
  journal={Nature},
  volume={518},
  number={7539},
  pages={371--375},
  year={2015},
  publisher={Nature Publishing Group}
}

@article{heusel2017gans,
  title={{GANs} trained by a two time-scale update rule converge to a local {N}ash equilibrium},
  author={Heusel, Martin and Ramsauer, Hubert and Unterthiner, Thomas and Nessler, Bernhard and Hochreiter, Sepp},
  journal={NeurIPS},
  volume={30},
  year={2017}
}

@article{frayer1985sexual,
  title={Sexual dimorphism},
  author={Frayer, David W and Wolpoff, Milford H},
  journal={Annu. Rev. Anthropol.},
  volume={14},
  pages={429--473},
  year={1985},
  publisher={JSTOR}
}

@inproceedings{lau2026efficient,
  title={Efficient score pre-computation for diffusion models via cross-matrix {Krylov} projection},
  author={Lau, Kaikwan and Na, Andrew S and Wan, Justin WL},
  booktitle={International Conference on Pattern Recognition},
  pages={697--712},
  year={2026},
  organization={Springer}
}

@inproceedings{kingma2014auto,
  title={Auto-encoding variational {B}ayes},
  author={Kingma, Diederik P and Welling, Max},
  booktitle={ICLR},
  year={2014}
}

@book{osher2004level,
  title={Level Set Methods and Dynamic Implicit Surfaces},
  author={Osher, Stanley and Fedkiw, Ronald},
  series={Applied Mathematical Sciences},
  volume={153},
  publisher={Springer},
  address={New York},
  year={2003}
}

@article{laird1982random,
  title={Random-effects models for longitudinal data},
  author={Laird, Nan M and Ware, James H},
  journal={Biometrics},
  volume={38},
  number={4},
  pages={963--974},
  year={1982}
}

@book{gelman2007data,
  title={Data Analysis Using Regression and Multilevel/Hierarchical Models},
  author={Gelman, Andrew and Hill, Jennifer},
  publisher={Cambridge University Press},
  address={Cambridge},
  year={2007}
}

@article{mantel1967detection,
  title={The detection of disease clustering and a generalized regression approach},
  author={Mantel, Nathan},
  journal={Cancer Res.},
  volume={27},
  number={2},
  pages={209--220},
  year={1967}
}

@article{mctavish2025complete,
  title={A complete and dynamic tree of birds},
  author={McTavish, Emily Jane and Gerbracht, Jeff A and Holder, Mark T and Iliff, Marshall J and Lepage, Denis and Rasmussen, Pamela C and Redelings, Benjamin D and S{\'a}nchez Reyes, Luna L and Miller, Eliot T},
  journal={Proc. Natl. Acad. Sci.},
  volume={122},
  number={18},
  pages={e2409658122},
  year={2025},
}

@article{dinnage2025generative,
  title={Generative {AI} extracts ecological meaning from the complex three dimensional shapes of bird bills},
  author={Dinnage, Russell and Kleineberg, Marian},
  journal={PLOS Comput. Biol.},
  volume={21},
  number={3},
  pages={e1012887},
  year={2025},
  doi={10.1371/journal.pcbi.1012887}
}

@inproceedings{khurana2024hierarchical,
  title={Hierarchical conditioning of diffusion models using tree-of-life for studying species evolution},
  author={Khurana, Mridul and Daw, Arka and Maruf, M and Uyeda, Josef C and Dahdul, Wasila and Charpentier, Caleb and Bak{\i}{\c{s}}, Yasin and Bart, Jr., Henry L and Mabee, Paula M and Lapp, Hilmar and Balhoff, James P and Chao, Wei-Lun and Stewart, Charles and Berger-Wolf, Tanya and Karpatne, Anuj},
  booktitle={ECCV},
  publisher={Springer},
  year={2024},
  volume={15147},
  series={Lecture Notes in Computer Science},
  pages={137--153}
}

@inproceedings{elhamod2023discovering,
  title={Discovering novel biological traits from images using phylogeny-guided neural networks},
  author={Elhamod, Mohannad and Khurana, Mridul and Manogaran, Harish Babu and Uyeda, Josef C and Balk, Meghan A and Dahdul, Wasila and Bak{\i}{\c{s}}, Yasin and Bart, Jr., Henry L and Mabee, Paula M and Lapp, Hilmar and Balhoff, James P and Charpentier, Caleb and Carlyn, David and Chao, Wei-Lun and Stewart, Charles V and Rubenstein, Daniel I and Berger-Wolf, Tanya and Karpatne, Anuj},
  booktitle={KDD},
  pages={3966--3978},
  year={2023}
}

@inproceedings{zhang2025cellflux,
  title={{CellFlux}: Simulating cellular morphology changes via flow matching},
  author={Zhang, Yuhui and Su, Yuchang and Wang, Chenyu and Li, Tianhong and Wefers, Zoe and Nirschl, Jeffrey J and Burgess, James and Ding, Daisy and Lozano, Alejandro and Lundberg, Emma and Yeung-Levy, Serena},
  booktitle={ICML},
  series={PMLR},
  volume={267},
  pages={76111--76138},
  year={2025}
}

@article{abdollahzadeh2025survey,
  title={A survey on generative modeling with limited data, few shots, and zero shot},
  author={Abdollahzadeh, Milad and Liu, Guimeng and Malekzadeh, Touba and Teo, Christopher T H and Chandrasegaran, Keshigeyan and Cheung, Ngai-Man},
  journal={Trans. Mach. Learn. Res.},
  year={2025}
}

@inproceedings{shim2023diffusion,
  title={Diffusion-based signed distance fields for {3D} shape generation},
  author={Shim, Jaehyeok and Kang, Changwoo and Joo, Kyungdon},
  booktitle={CVPR},
  pages={20887--20897},
  year={2023}
}

@article{lau2025robust,
  title = {Robust parametric estimation of avian cranial morphology},
  volume = {22},
  ISSN = {1553-7358},
  DOI = {10.1371/journal.pcbi.1014941},
  number = {11},
  journal = {PLOS Comput. Biol.},
  publisher = {Public Library of Science (PLoS)},
  author = {Lau,  Kaikwan and Choi,  Gary P. T.},
  year = {2026},
  pages = {e1014941}
}

@inproceedings{loshchilov2019decoupled,
  title={Decoupled weight decay regularization},
  author={Loshchilov, Ilya and Hutter, Frank},
  booktitle={ICLR},
  year={2019}
}

@article{li2023progressive,
  title={Progressive generation of {3D} point clouds with hierarchical consistency},
  author={Li, Peipei and Liu, Xiyan and Huang, Jizhou and Xia, Deguo and Yang, Jianzhong and Lu, Zhen},
  journal={Pattern Recognit.},
  volume={136},
  pages={109200},
  year={2023},
}

@article{yang2026unified,
  title={A unified multi-modality conditional latent diffusion model for point cloud generation},
  author={Yang, Yihang and Zhao, Zibo and Yin, Fukun and Liu, Wen and Ding, Yuhan and Jiang, Biao and Yu, Gang and Chen, Tao},
  journal={Pattern Recognit.},
  volume={172},
  pages={112716},
  year={2026},
}

@article{zhu2024advancements,
  title={Advancements in point cloud data augmentation for deep learning: A survey},
  author={Zhu, Qinfeng and Fan, Lei and Weng, Ningxin},
  journal={Pattern Recognit.},
  volume={153},
  pages={110532},
  year={2024},
}

@inproceedings{liu2025g2pdiffusion,
  title={{G2PDiffusion}: Cross-species genotype-to-phenotype prediction via evolutionary diffusion},
  author={Liu, Mengdi and Gao, Zhangyang and Chang, Hong and Li, Stan Z. and Shan, Shiguang and Chen, Xilin},
  booktitle={ICCV},
  pages={20705--20714},
  year={2025},
}

\newpage
\centerline{\Large\textbf{Supplementary Information}}
\appendix
\renewcommand\thefigure{S\arabic{figure}}    
\setcounter{figure}{0}
\renewcommand\thetable{S\arabic{table}}    
\setcounter{table}{0}
\renewcommand{\thesection}{S\arabic{section}}

\vspace{1cm}

\noindent{The sections in this supplementary document follow the order of the main text. Residual CFM is also called the pooled residual flow or the flow, and the GMM baseline is the per-species Gaussian. Sections~\ref{sec:si_impl}--\ref{sec:si_eval} give the implementation, distance matrices, and evaluator. Sections~\ref{sec:si_tree_sampler}--\ref{sec:si_anat} give the details of each study, and Section~\ref{sec:si_historical} presents supplementary analyses with conditional CFM. The mathematical background is in main-text Section~3.}

\section{Algorithm and Implementation Details}
\label{sec:si_impl}
{Main-text Fig.~2 outlines the two stages. The experiments use the same collection of 100 scans, per-skull normalization (centred on the vertex mean, largest absolute coordinate 1) and geometric evaluator (Section~\ref{sec:si_eval}); historical protocols are identified in Section~\ref{sec:si_historical}. Algorithm~\ref{alg:generation} gives generation with the final model, and Table~\ref{stab:arch} lists the network architectures. The primary sampler comparison uses a frozen tree-trained representation (Section~\ref{sec:si_tree_sampler}); the proxy-based comparison, component studies and morphometric audit use a separate representation and the protocols below.}

\begin{algorithm}[h!]
\caption{Residual CFM Generation}
\label{alg:generation}
\noindent\textbf{Require:} species $s$, centroid $\bmu_s$, velocity network $v_\theta$, residual scale $\sigma_r$, temperature $T$ ($T=1$ unless stated), Euler steps $K=50$
\begin{enumerate}
\setlength{\itemsep}{0pt}\setlength{\parsep}{0pt}\setlength{\topsep}{2pt}
\item $\bx\sim\cN(\mathbf{0},T^2\mathbf{I}_{64})$
\item \textbf{for} $k=0,\ldots,K-1$ \textbf{do}
\item \hspace*{1em}$\bx\leftarrow\bx+\tfrac{1}{K}\,v_\theta(\bx,\,k/K)$\\
\hspace*{1em}\textit{Neutral inputs: null species token, zero size, zero phylogeny.}
\item \textbf{end for}
\item $\bz_{\mathrm{gen}}\leftarrow\bmu_s+\sigma_r\,\bx$
\item evaluate $f_\theta(\bz_{\mathrm{gen}},\cdot)$ on a $256^3$ grid over $[-1,1]^3$ (spacing $2/255$) and extract the zero level set by Marching Cubes~\cite{lorensen1998marching}
\item \textbf{return} mesh as extracted (it may be open where it meets the grid boundary)
\end{enumerate}
\end{algorithm}

\paragraph{{Representation (Stage 1).}} {The decoder $f_\theta$ and the latent codes are trained jointly with two Adam optimizers~\cite{kingma2014adam} at default momentum, with learning rates $5\times10^{-4}$ for the decoder and $10^{-3}$ for the codes, both halved every 500 epochs. Training runs for 2,000 epochs with batches of 16,384 SDF samples drawn from a fixed random subset of 50,000 of the 100,000 precomputed samples per skull, half near the surface and half uniform in $[-1,1]^3$. Near-surface samples perturb surface points by Gaussian noise with standard deviation 0.003 for 70\% of them and 0.03 for the rest. Before sampling, each mesh is centred on its vertex mean and scaled so that its largest absolute coordinate is 1, which removes absolute size. The latent weight is $\alpha=10^{-3}$ and the distance-loss weight $\beta=5\times10^{-3}$. The distance loss, the negative Pearson correlation over all $N(N-1)/2$ specimen pairs, is added every tenth batch. Codes are initialized from $\cN(0,0.01^2)$, and we use the final checkpoint (epoch 2,000). A proxy-based sensitivity check uses the checkpoint with the lowest training loss (Table~\ref{stab:best}), whose provenance records keep a legacy filename.}

\paragraph{{Final flow (Stage 2).}} {PhyloFlowNet (Table~\ref{stab:arch}) receives neutral inputs: a null species token, zero size and a zero phylogenetic embedding. It is trained on the 100 real residuals, standardized by their scalar SD, for 10,000 updates of 100 residuals drawn uniformly with replacement. Pooling shares information about variation across species; it does not add independent observations, and residuals share estimated species centroids. We use AdamW~\cite{loshchilov2019decoupled} with initial learning rate $10^{-4}$, weight decay $10^{-5}$ and a cosine schedule reaching $10^{-6}$, gradient clipping at norm 1, variance weight $\lambda_v=0.1$ and dropout 0.05. Flows are trained with seeds 42, 43 and 44 and evaluated at their final checkpoints with $T=1$, 50 Euler steps and no guidance. The proxy-based and dated-tree sampler comparisons share this schedule but use different representations, baseline definitions and sampling runs.}

\begin{table}[t]
\centering
\footnotesize
\resizebox{\ifdim\width>\linewidth\linewidth\else\width\fi}{!}{%
{\color{black}
\begin{tabular}{llr}
\toprule
\multicolumn{3}{l}{\textit{(a) DeepSDF decoder ($D=64$)}}\\
Layer & Dimensions, notes & Parameters\\
\midrule
Linear 1 & 64+3=67 $\to$ 512; weight norm, ReLU & 35,328\\
Linear 2--4 & 512 $\to$ 512; weight norm, ReLU & 789,504\\
Linear 5 & 512+67=579 $\to$ 512; skip input, weight norm, ReLU & 297,472\\
Linear 6--7 & 512 $\to$ 512; weight norm, ReLU & 526,336\\
Linear 8 & 512 $\to$ 1; tanh output & 513\\
\textbf{Total} & & \textbf{1,649,153}\\
\midrule
\multicolumn{3}{l}{\textit{(b) PhyloFlowNet velocity network}}\\
Component & Dimensions, notes & Parameters\\
\midrule
Time embedding & sinusoidal 128 $\to$ 128 $\to$ 128 & 33,024\\
Species embedding & 25 $\times$ 32 (24 species + null) & 800\\
Size projection & 3 $\to$ 32 $\to$ 32 & 1,184\\
Phylogeny projection & 32 $\to$ 64 $\to$ 32 & 4,192\\
Input projection & (64+128+96) $\to$ 128 & 36,992\\
Residual blocks ($\times3$) & LayerNorm, 128 $\to$ 256 $\to$ 128, GELU, dropout 0.05 & 198,528\\
Output & LayerNorm, 128 $\to$ 64 & 8,512\\
\textbf{Total} & & \textbf{283,232}\\
\bottomrule
\end{tabular}
}%
}
\caption{{Network architectures. In the final model the species, size and phylogeny inputs of PhyloFlowNet receive constant neutral values (null token, zeros) and act as fixed biases. The latent library contains 100 codes of dimension 64.}}
\label{stab:arch}
\end{table}

\paragraph{{Neutral conditioning.}} {Constant species, size and phylogeny inputs remove direct conditioning of the velocity network. Distance supervision still enters through the representation, because the decoder and codes are optimized with the supplied matrix and the centroids and residuals are computed from those codes. Removing only the flow's phylogenetic input therefore tests direct conditioning, whereas the controls of Table~\ref{stab:phylo_controls} retrain the representation to test supervision itself.}

\paragraph{{Variance penalty and temperature.}} {The penalty is
\begin{equation}
\mathcal{L}_\text{var} = \max\!\left(0,\;\mathrm{Var}(\mathbf{x}_1) - \mathrm{Var}(\hat{\mathbf{x}}_1)\right),\qquad
\hat{\mathbf{x}}_1 = \mathbf{x}_{0.5} + 0.5\,v_\theta(\mathbf{x}_{0.5}, 0.5),
\label{eq:lvar}
\end{equation}
where, for a minibatch $X\in\mathbb{R}^{B\times 64}$, $\operatorname{Var}(X)=\frac{1}{64B-1}\sum_{b,d}(X_{bd}-\overline X)^2$ is a single scalar over all entries, with $\overline X$ their mean, and the total loss is $\mathcal{L}_\text{CFM}+0.1\,\mathcal{L}_\text{var}$.}

{The variance term penalizes a deficit in the scalar unbiased variance of a midpoint-based one-step estimate (Eq.~\eqref{eq:lvar}). The midpoint contains a training target, so this estimate differs from the endpoint of the 50-step trajectory sampled from noise, and the penalty cannot guarantee the variance or covariance of generated residuals. Removing the penalty had no detectable effect under the final-model ablation protocol (Table~\ref{stab:s03}; main-text Section~5.6), so these results do not establish a diversity benefit.}

{Temperature calibration is a separate sampling procedure. It selects $T$ by 30 bisection iterations on $[0.5,4]$, using 4,000 fixed noise draws, so that the scalar variance of the generated standardized residuals matches that of the training residuals. Both variances pool all sample and coordinate entries with denominator $BD-1$ ($B=4{,}000$ generated or $B=100$ training residuals, $D=64$). At $T=1$, the ratio of generated to training variance is 0.37--0.43 under the ablation schedule. This is a variance ratio, not a ratio of standard deviations, and matching it on 4,000 draws guarantees neither the same variance on new draws nor the preservation of mesh diversity.}

\paragraph{{Ablation schedule.}} {The component studies (Table~\ref{stab:s03} and main-text Fig.~7b) train the final model and its variants for 10,000 updates with batches of 256 drawn with replacement, using the optimizer settings above. Seeds 42--44 use identical initial weights across variants. This schedule differs from the sampler comparisons in batch size, species schedule and sampling noise. Species-conditioned variants receive the species token, the normalized size of the scheduled species and the 32-dimensional phylogenetic embedding, and with probability 0.1 the species token is replaced by the null token for the whole batch. Sampling uses no guidance.}

\section{{Species Distance Matrices}}
\label{sec:si_matrices}
\begin{figure}[t!]
    \centering
    \makebox[\linewidth][c]{\includegraphics[width=1\linewidth]{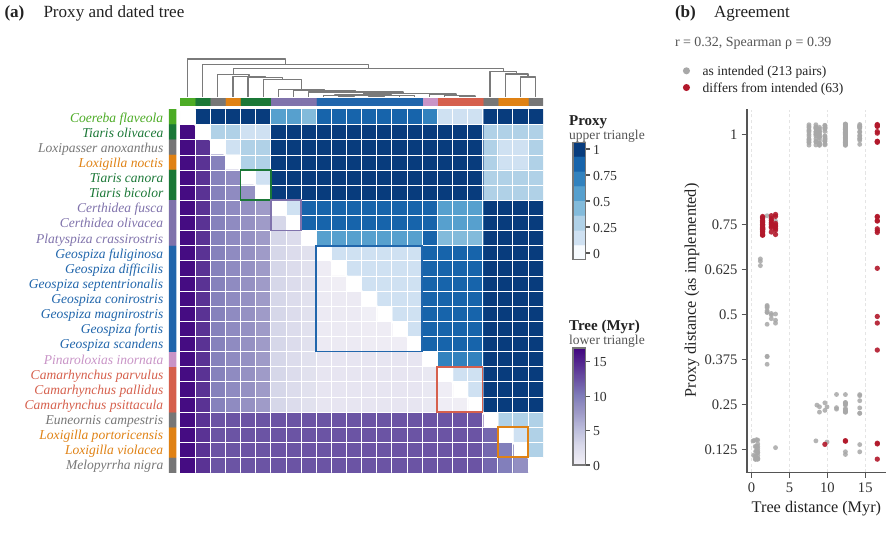}}
    \caption{{\textbf{The two species distance matrices.} (a)~Taxonomic proxy as implemented (upper triangle) and patristic distances on the dated summary tree (lower triangle, Myr), ordered by average-linkage clustering of the tree, with genera outlined. (b)~Agreement over the 276 species pairs ($r=0.32$, Spearman 0.39). Red points are the 63 pairs whose implemented proxy value differs from the intended scheme.}}
    \label{sfig:matrices}
\end{figure}

{Main-text Fig.~4 illustrates how supplied species distances enter the consistency loss, using the implemented proxy. Figure~\ref{sfig:matrices} compares this proxy with the dated-tree distances and identifies the 63 species pairs affected by the implementation errors.}

\paragraph{{Taxonomic proxy.}} {The original code built the proxy from a hand-set table of genus-pair distances: 0.5 within a genus, 1.5--3.0 between Darwin's finch genera, 1.0 between other genera and 4.0 between Darwin's finches and other genera, all scaled by 1/4. It inferred the genus from the abbreviated species name. Two implementation errors affect 63 of the 276 species pairs (Table~\ref{stab:proxy}). The lookup sorts keys alphabetically, so three genus pairs stored in reverse order receive the default distance of 3.0 (0.75 after scaling), and the parser reads \textit{Coereba} (`C.') as \textit{Camarhynchus} and \textit{Loxipasser} (`L.') as \textit{Loxigilla}. The original function reproduces the stored training matrix exactly. The proxy-based sampler comparison, component studies and morphometric audit keep this representation.}

\begin{table}[t]
\centering
\footnotesize
\resizebox{\ifdim\width>\linewidth\linewidth\else\width\fi}{!}{%
{\color{black}
\begin{tabular}{llccc}
\toprule
Genus pair & & Intended & As trained & Species pairs\\
\midrule
\textit{Camarhynchus} & \textit{Geospiza} & 0.375 & 0.750 & 21\\
\textit{Certhidea} & \textit{Geospiza} & 0.625 & 0.750 & 14\\
\textit{Coereba} & \textit{Geospiza} & 1.000 & 0.750 & 7\\
\textit{Camarhynchus} & \textit{Coereba} & 1.000 & 0.125 & 3\\
\textit{Coereba} & \textit{Loxigilla} & 0.250 & 1.000 & 3\\
\textit{Coereba} & \textit{Tiaris} & 0.250 & 1.000 & 3\\
\textit{Loxigilla} & \textit{Loxipasser} & 0.250 & 0.125 & 3\\
\textit{Certhidea} & \textit{Coereba} & 1.000 & 0.500 & 2\\
\textit{Certhidea} & \textit{Platyspiza} & 0.625 & 0.750 & 2\\
\textit{Coereba} & \textit{Euneornis} & 0.250 & 1.000 & 1\\
\textit{Coereba} & \textit{Loxipasser} & 0.250 & 1.000 & 1\\
\textit{Coereba} & \textit{Melopyrrha} & 0.250 & 1.000 & 1\\
\textit{Coereba} & \textit{Pinaroloxias} & 1.000 & 0.625 & 1\\
\textit{Coereba} & \textit{Platyspiza} & 1.000 & 0.375 & 1\\
\midrule
All differing pairs & & & & 63 of 276\\
\bottomrule
\end{tabular}
}%
}
\caption{{Taxonomic proxy: values intended by the genus table of the original code and values used in training, after scaling to $[0,1]$. Differences arise from an order-dependent genus-pair lookup, in which pairs missing from the table fall to the default 0.75, and from reading the genus from the abbreviated name (\textit{Coereba} as \textit{Camarhynchus}, \textit{Loxipasser} as \textit{Loxigilla}). The as-trained values reproduce the stored training matrix exactly.}}
\label{stab:proxy}
\end{table}

\paragraph{{Dated tree.}} {The tree and dataset composition are shown in main-text Fig.~3. We reproduced the archived tree-distance matrix from the maximum-clade-credibility summary tree distributed as \path{Tree_versions/Aves_1.5/Clements2024/summary_dated_clements.nex} in the McTavish et al.\ data repository (\url{https://github.com/McTavishLab/AvesData})~\cite{mctavish2025complete}. Table~\ref{stab:tipmap} maps the dataset identifiers to the 24 tree tips, including spelling and genus-name changes. Patristic distances sum branch lengths along the path between two species, in Myr. The reconstruction reproduces all 276 entries embedded in the experiment scripts to their stored precision of six decimal places (range 0.192978--16.586416\,Myr). The matrix is divided by its maximum before use in the regularizer and the centroid predictor. The inspected records do not keep the original download receipt or extraction script. McTavish et al.\ report that nearly 40\% of internal-node dates in their synthesis are interpolated, so these distances are not exact divergence times. The proxy and tree matrices correlate at $r=0.32$ (Spearman 0.39) over species pairs (Fig.~\ref{sfig:matrices}).}

\begin{table}[t]
{
\centering
\footnotesize
\setstretch{1.1}
{\color{black}
\begin{tabular}{ll}
\toprule
Dataset identifier & Tip label in the released tree\\
\midrule
\texttt{C.Pallidus} & \texttt{Camarhynchus\_pallidus}\\
\texttt{C.Parvulus} & \texttt{Camarhynchus\_parvulus}\\
\texttt{C.Psittacula} & \texttt{Camarhynchus\_psittacula}\\
\texttt{C.flaveola} & \texttt{Coereba\_flaveola}\\
\texttt{C2.Fusca} & \texttt{Certhidea\_fusca}\\
\texttt{C2.Olivacea} & \texttt{Certhidea\_olivacea}\\
\texttt{E.campestris} & \texttt{Euneornis\_campestris}\\
\texttt{G.Conirostris} & \texttt{Geospiza\_conirostris}\\
\texttt{G.Difficilis} & \texttt{Geospiza\_difficilis}\\
\texttt{G.Fortis} & \texttt{Geospiza\_fortis}\\
\texttt{G.Fuliginosa} & \texttt{Geospiza\_fuliginosa}\\
\texttt{G.Magnirostris} & \texttt{Geospiza\_magnirostris}\\
\texttt{G.Scandens} & \texttt{Geospiza\_scandens}\\
\texttt{G.Septentrionalist} & \texttt{Geospiza\_septentrionalis}\\
\texttt{L.anoxanthus} & \texttt{Loxipasser\_anoxanthus}\\
\texttt{L.noctis} & \texttt{Loxigilla\_noctis}\\
\texttt{L.portoricensis} & \texttt{Melopyrrha\_portoricensis}\\
\texttt{L.violacea} & \texttt{Melopyrrha\_violacea}\\
\texttt{M.nigra} & \texttt{Melopyrrha\_nigra}\\
\texttt{P.Crassirostris} & \texttt{Platyspiza\_crassirostris}\\
\texttt{P2.Inornata} & \texttt{Pinaroloxias\_inornata}\\
\texttt{T.Bicolor} & \texttt{Melanospiza\_bicolor}\\
\texttt{T.canora} & \texttt{Phonipara\_canora}\\
\texttt{T.olivacea} & \texttt{Tiaris\_olivaceus}\\
\bottomrule
\end{tabular}
\par
}}
\caption{{Mapping used to reconstruct the archived dated-tree matrix from the Aves\_1.5, Clements 2024 maximum-clade-credibility summary tree. Dataset identifiers keep their original capitalization and spelling, and the second column gives the exact tree labels and changed genus names. The reconstruction reproduces the stored matrix but does not establish museum specimen identities.}}
\label{stab:tipmap}
\end{table}

\paragraph{{Mantel tests.}} {Agreement between a representation and a matrix is the Pearson correlation between the 276 pairwise distances among the 24 species centroids and the corresponding matrix entries. The test permutes species labels 9,999 times and reports the upper-tail probability $p=(1+\#\{r_{\mathrm{perm}}\ge r_{\mathrm{obs}}\})/10{,}000$~\cite{mantel1967detection}, a one-sided test of positive association. Applied to the matrix used in training, the correlation and its $p$-value describe target fit rather than independent phylogenetic structure.}

\paragraph{{Shuffled matrices.}} {For the supervision controls, we selected three derangements of the proxy's species labels whose correlations with both the proxy and the tree satisfy $|r|<0.1$ ($r=-0.03$ to $-0.06$ against the proxy and $-0.04$ to $0.02$ against the tree), which makes them negative controls rather than a null sample. For the withheld-species test, we drew three random permutations of the tree's species labels, whose correlations with the tree are $-0.19$, $0.21$ and $-0.24$.}

================================================================
\section{{Evaluation Protocol}}
\label{sec:si_eval}
\paragraph{{Evaluator.}} {All geometric comparisons use the same frozen evaluator. Meshes are compared in the decoder frame, with real skulls mapped into it by their saved centroid and scale. Each mesh is sampled at 10,000 area-weighted surface points with two fixed seeds. For each seed, pairwise Chamfer distances follow main-text Eq.~(7), and nn-CD, MMD-CD, COV and diversity are computed from that seed's distance matrix. The two metric values are then averaged, so averaging follows nearest-neighbour assignment rather than preceding it. MMD-CD gives each reference species equal weight and each specimen equal weight within its species. For reference sets $\mathcal R_s$ and generated set $\mathcal G$, it is $\frac{1}{S}\sum_{s=1}^{S}\frac{1}{|\mathcal R_s|}\sum_{r\in\mathcal R_s}\min_{g\in\mathcal G}\mathrm{CD}(r,g)$. The MorphFID reference statistics use the same species-balanced weighting. COV assigns each generated skull to its nearest real skull without a threshold. MorphFID descriptors are z-scored with species-balanced statistics of the 100 real skulls, and non-watertight meshes use convex-hull volume.}

\paragraph{{Cohorts and statistics.}} {The dated-tree comparison and the final-model ablations make 140 attempts per sampler and seed, namely 100 species-mixture skulls and ten for each of four focal species, and every attempt counts. The mixture-only EM Gaussian mixture of Table~\ref{stab:s02} makes 100 attempts, and withheld-species generation makes 40 per centroid predictor. 1-NNA is computed separately for each surface seed and then averaged. It describes separation across the whole mixture, not anatomical acceptability.}

\paragraph{{Reference values.}} {The copying reference decodes the training code of a randomly selected specimen of the scheduled species (1,000 draws) on the proxy-based decoder and gives nn-CD 0.25. For the tree-trained decoder, the 16 empirical draws that reuse a training code give 0.234 on average (Section~\ref{sec:si_tree_sampler}). The real-skull reference compares each skull with the other 99 and gives 1.467 under the dated-tree evaluator (1.4646 and 1.4691 for the two surface seeds). No equivalence margin was specified, so nonsignificant differences do not establish equivalence.}

\section{{Dated-Tree Sampler Comparison}}
\label{sec:si_tree_sampler}
\paragraph{{Representation.}} {This comparison supplies main-text Table~1. The archived dated-tree representation was trained on all 100 skulls from 24 species for 2,000 epochs with seed 42, and all four samplers share its frozen decoder and fitted codes without retraining. A preflight check confirmed specimen and species ordering, equality of stored and native checkpoint codes, decoder arithmetic parity and the replay of an archived Chamfer calculation. The tree LPC of 0.99137 measures fit to the training target.}

\paragraph{{Samplers.}} {Let $\bmu_s$ be the mean fitted code of species $s$, $\br_i=\bz_i-\bmu_{s(i)}$, $\overline{\br}=N^{-1}\sum_i\br_i$ and $N=100$. The pooled residual flow uses the Stage-2 schedule of Section~\ref{sec:si_impl}, with separate flows for seeds 42, 43 and 44. The pooled Gaussian generates
\[
\bz=\bmu_s+\overline{\br}+L\boldsymbol\epsilon,
\qquad
\boldsymbol\epsilon\sim\mathcal N(\mathbf0,\mathbf I),\quad
LL^\top=\frac1N\sum_{i=1}^N(\br_i-\overline{\br})(\br_i-\overline{\br})^\top,
\]
with the full empirical population covariance, no diagonal jitter and $L$ its symmetric positive-semidefinite square root. The original pooled-Gaussian code could not be recovered, so this fit is specified here. Empirical sampling draws uniformly with replacement from the same 100 residuals and adds the draw to $\bmu_s$. The per-species Gaussian, the GMM baseline of the main text, uses the archived \texttt{PhyloGMM} shrinkage implementation. Its global covariance is the unbiased full-code covariance plus $10^{-4}\mathbf I$, species with at least three specimens use shrinkage weight $2/(n_s+2)$, and smaller species use half the global covariance. Sampling uses a symmetric covariance square root, which preserves the specified distributions but not the original random draws.}

\paragraph{{Matched slots and evaluation.}} {Every sampler and seed uses the same 140 ordered slots: ten skulls for each of the four focal species, followed by 100 mixture slots that cycle through the 24 species indices, so the mixture is approximately species-uniform. Flow and Gaussian arms share CPU standard-normal noise with seed $s+100000$, and empirical donor indices use NumPy's \texttt{default\_rng} with seed $s+500000$. Decoder evaluation is FP32 with TF32 disabled, and meshes are extracted on $256^3$ points over $[-1,1]^3$ without repair or component filtering. MorphFID-8 is distinct from the 12-measure audit of Section~\ref{sec:si_anat}.}

\begin{table}[t]
\centering\footnotesize
\setlength{\tabcolsep}{3.2pt}
\resizebox{\ifdim\width>\linewidth\linewidth\else\width\fi}{!}{%
{\color{black}
\begin{tabular}{llrrrrrr}
\toprule
Sampler & Seed & nn-CD$\downarrow$ & MMD-CD$\downarrow$ & COV$\uparrow$ & 1-NNA$\to0.5$ & Diversity & MorphFID-8$\downarrow$\\
\midrule
Residual flow & 42 & 1.248 & 1.181 & 0.530 & 0.532 & 10.350 & 1.436\\
Residual flow & 43 & 1.187 & 1.171 & 0.500 & 0.568 & 10.649 & 1.925\\
Residual flow & 44 & 1.174 & 1.173 & 0.510 & 0.550 & 10.370 & 1.661\\
\textbf{Mean $\pm$ SD} & 42--44 & 1.203\,$\pm$\,0.039 & 1.175\,$\pm$\,0.005 & 0.513\,$\pm$\,0.015 & 0.550\,$\pm$\,0.018 & 10.456\,$\pm$\,0.167 & 1.674\,$\pm$\,0.244\\\addlinespace
Pooled Gaussian & 42 & 1.301 & 1.215 & 0.600 & 0.512 & 10.227 & 0.952\\
Pooled Gaussian & 43 & 1.322 & 1.298 & 0.530 & 0.565 & 11.049 & 1.463\\
Pooled Gaussian & 44 & 1.360 & 1.224 & 0.500 & 0.522 & 10.600 & 0.892\\
\textbf{Mean $\pm$ SD} & 42--44 & 1.328\,$\pm$\,0.030 & 1.246\,$\pm$\,0.045 & 0.543\,$\pm$\,0.051 & 0.533\,$\pm$\,0.028 & 10.625\,$\pm$\,0.412 & 1.102\,$\pm$\,0.314\\\addlinespace
Empirical residual & 42 & 1.380 & 1.207 & 0.540 & 0.522 & 11.064 & 0.803\\
Empirical residual & 43 & 1.341 & 1.132 & 0.560 & 0.480 & 11.895 & 0.930\\
Empirical residual & 44 & 1.421 & 1.256 & 0.495 & 0.555 & 10.729 & 0.814\\
\textbf{Mean $\pm$ SD} & 42--44 & 1.381\,$\pm$\,0.040 & 1.198\,$\pm$\,0.063 & 0.532\,$\pm$\,0.033 & 0.519\,$\pm$\,0.038 & 11.229\,$\pm$\,0.600 & 0.849\,$\pm$\,0.070\\\addlinespace
Per-species Gaussian & 42 & 1.453 & 1.319 & 0.585 & 0.400 & 11.584 & 0.786\\
Per-species Gaussian & 43 & 1.605 & 1.360 & 0.605 & 0.482 & 12.850 & 0.932\\
Per-species Gaussian & 44 & 1.687 & 1.544 & 0.590 & 0.538 & 13.502 & 1.203\\
\textbf{Mean $\pm$ SD} & 42--44 & 1.582\,$\pm$\,0.119 & 1.408\,$\pm$\,0.120 & 0.593\,$\pm$\,0.010 & 0.473\,$\pm$\,0.069 & 12.645\,$\pm$\,0.975 & 0.973\,$\pm$\,0.211\\\addlinespace
\bottomrule
\end{tabular}
}}
\caption{{Dated-tree species-mixture comparison: 100 generated meshes per sampler and seed against the same 100 real references (nn-CD, MMD-CD and pairwise diversity $\times10^{-3}$, COV and 1-NNA as fractions). Means and sample SD use three generator seeds after averaging the two surface-seed values. All 100 attempts per row were extracted and evaluated, including open surfaces.}}
\label{stab:tree_mixture}
\end{table}

\begin{table}[t]
\centering\footnotesize
\resizebox{\ifdim\width>\linewidth\linewidth\else\width\fi}{!}{%
{\color{black}
\begin{tabular}{lcccc}
\toprule
Sampler & \textit{C. parvulus}$\to100\%$ & \textit{G. fortis}$\to100\%$ & \textit{G. fuliginosa}$\to100\%$ & \textit{G. magnirostris}$\to100\%$\\
\midrule
Residual flow & 46.7\,$\pm$\,14.2 & 55.0\,$\pm$\,7.4 & 60.9\,$\pm$\,9.5 & 50.9\,$\pm$\,9.6\\
Pooled Gaussian & 45.0\,$\pm$\,18.0 & 56.1\,$\pm$\,3.0 & 60.9\,$\pm$\,13.4 & 72.2\,$\pm$\,15.0\\
Empirical residual & 50.2\,$\pm$\,12.4 & 63.7\,$\pm$\,5.4 & 76.4\,$\pm$\,4.4 & 60.5\,$\pm$\,16.2\\
Per-species Gaussian & 79.5\,$\pm$\,19.4 & 87.0\,$\pm$\,3.1 & 80.9\,$\pm$\,25.3 & 100.1\,$\pm$\,3.2\\
\bottomrule
\end{tabular}
}}
\caption{{Dated-tree within-species mesh diversity as a percentage of the real species' mean pairwise CD (mean $\pm$ sample SD over three generator seeds, ten skulls per focal species and seed). The real reference distances are 9.735, 7.347, 7.041 and 6.133 ($\times10^{-3}$). The percentages describe retained observed variation in this collection, not population variance.}}
\label{stab:tree_diversity}
\end{table}

\begin{table}[t]
\centering\footnotesize
{\color{black}
\begin{tabular}{lccc}\toprule
Sampler & nn-CD$\downarrow$ & MMD-CD$\downarrow$ & COV$\uparrow$\\\midrule
\multicolumn{4}{l}{\textit{C.~parvulus}}\\
Residual flow & 1.860\,$\pm$\, 0.201 & 1.655\,$\pm$\, 0.449 & 0.767\,$\pm$\, 0.058\\
Pooled Gaussian & 1.790\,$\pm$\, 0.233 & 1.573\,$\pm$\, 0.313 & 0.767\,$\pm$\, 0.058\\
Empirical residual & 2.292\,$\pm$\, 0.106 & 1.723\,$\pm$\, 0.267 & 0.700\,$\pm$\, 0.100\\
Per-species Gaussian & 1.959\,$\pm$\, 0.197 & 1.223\,$\pm$\, 0.266 & 0.767\,$\pm$\, 0.058\\
\addlinespace
\multicolumn{4}{l}{\textit{G.~fortis}}\\
Residual flow & 1.674\,$\pm$\, 0.521 & 1.071\,$\pm$\, 0.206 & 0.600\,$\pm$\, 0.200\\
Pooled Gaussian & 2.013\,$\pm$\, 0.192 & 1.453\,$\pm$\, 0.181 & 0.600\,$\pm$\, 0.200\\
Empirical residual & 2.071\,$\pm$\, 0.149 & 1.261\,$\pm$\, 0.223 & 0.633\,$\pm$\, 0.153\\
Per-species Gaussian & 2.035\,$\pm$\, 0.513 & 1.445\,$\pm$\, 0.286 & 0.867\,$\pm$\, 0.115\\
\addlinespace
\multicolumn{4}{l}{\textit{G.~fuliginosa}}\\
Residual flow & 1.989\,$\pm$\, 0.240 & 1.564\,$\pm$\, 0.191 & 0.833\,$\pm$\, 0.000\\
Pooled Gaussian & 1.804\,$\pm$\, 0.481 & 1.737\,$\pm$\, 0.760 & 0.722\,$\pm$\, 0.192\\
Empirical residual & 1.542\,$\pm$\, 0.156 & 1.354\,$\pm$\, 0.143 & 0.722\,$\pm$\, 0.096\\
Per-species Gaussian & 1.825\,$\pm$\, 0.143 & 1.384\,$\pm$\, 0.466 & 0.722\,$\pm$\, 0.192\\
\addlinespace
\multicolumn{4}{l}{\textit{G.~magnirostris}}\\
Residual flow & 1.237\,$\pm$\, 0.198 & 1.684\,$\pm$\, 0.075 & 0.600\,$\pm$\, 0.200\\
Pooled Gaussian & 1.458\,$\pm$\, 0.069 & 1.470\,$\pm$\, 0.155 & 0.733\,$\pm$\, 0.231\\
Empirical residual & 1.463\,$\pm$\, 0.264 & 1.696\,$\pm$\, 0.253 & 0.600\,$\pm$\, 0.200\\
Per-species Gaussian & 1.994\,$\pm$\, 0.390 & 1.465\,$\pm$\, 0.318 & 0.867\,$\pm$\, 0.115\\
\addlinespace
\bottomrule\end{tabular}
}
\caption{{Dated-tree focal-species fidelity and coverage (mean $\pm$ sample SD over seeds 42--44, ten skulls per species and seed, Chamfer values $\times10^{-3}$). The references contain 5, 5, 6 and 5 skulls. These are comparisons against training references.}}
\label{stab:tree_focal}
\end{table}

\paragraph{{Aggregation and 1-NNA.}} {Table~\ref{stab:tree_mixture} reports the individual mixture runs with their means and sample SD, and Tables~\ref{stab:tree_diversity} and~\ref{stab:tree_focal} report focal-species diversity, fidelity and coverage. 1-NNA uses the union of 100 real and 100 generated mixture meshes, excluding each sample itself, with ties resolved by the first locked union index (reference meshes first). It is not computed for the small focal cohorts. To read 1-NNA against chance, we relabelled the 200 meshes of each run at random 4,000 times, recomputed 1-NNA with the same neighbours and averaging, and took the 2.5 and 97.5 percentiles. The interval is about 0.41--0.58 in every run. The flow lies inside it in all three seeds (0.532, 0.568 and 0.550), as do the pooled Gaussian and empirical residual samplers, whereas the per-species Gaussian falls below it in seed 42 (0.400). The three generator seeds measure variation in flow training or baseline draws for a fixed representation, and the two surface seeds measure sensitivity to surface sampling. Apart from these intervals, no hypothesis tests were run on this comparison, and the proxy-based $p$-values of Table~\ref{stab:paired} do not apply to it.}

\begin{table}[t]
\centering\footnotesize
\setlength{\tabcolsep}{4pt}
\resizebox{\ifdim\width>\linewidth\linewidth\else\width\fi}{!}{%
{\color{black}
\begin{tabular}{lcccc}
\toprule
Sampler & Non-watertight, all$\downarrow$ & Boundary contact & Multiple components & Non-watertight, mixture$\downarrow$\\
\midrule
Residual flow & 198/420 (47.1\%) & 198 & 182 & 142/300\\
Pooled Gaussian & 165/420 (39.3\%) & 164 & 189 & 113/300\\
Empirical residual & 153/420 (36.4\%) & 153 & 189 & 94/300\\
Per-species Gaussian & 108/420 (25.7\%) & 108 & 182 & 76/300\\
\midrule
All four samplers & 624/1,680 (37.1\%) & 623 & 742 & 425/1,200\\
\bottomrule
\end{tabular}
}}
\caption{{Dated-tree topology across all three seeds. Each sampler has 420 outputs (300 mixture and 120 focal-species meshes), all extracted and evaluated. The boundary-contact and multi-component columns count all 420 outputs, and the last column counts the 300 mixture outputs. Every non-watertight mesh uses convex-hull volume, without repair. The 100 real meshes are watertight, all touch the normalized boundary and 27 have several components.}}
\label{stab:tree_topology}
\end{table}

\begin{table}[t]
\centering\footnotesize
\resizebox{\ifdim\width>\linewidth\linewidth\else\width\fi}{!}{%
{\color{black}
\begin{tabular}{lcccc}\toprule
Sampler & Original 8$\downarrow$ & Six descriptors$\downarrow$ & Closed subset, 8$\downarrow$ & Closed $n$, seeds 42/43/44\\\midrule
Residual flow & 1.674\,$\pm$\, 0.244 & 0.049\,$\pm$\, 0.006 & 0.559\,$\pm$\, 0.183 & 55/49/54\\
Pooled Gaussian & 1.102\,$\pm$\, 0.314 & 0.088\,$\pm$\, 0.023 & 0.610\,$\pm$\, 0.420 & 64/59/64\\
Empirical residual & 0.849\,$\pm$\, 0.070 & 0.077\,$\pm$\, 0.029 & 0.540\,$\pm$\, 0.226 & 70/67/69\\
Per-species Gaussian & 0.973\,$\pm$\, 0.211 & 0.389\,$\pm$\, 0.160 & 0.958\,$\pm$\, 0.149 & 73/80/71\\
\bottomrule\end{tabular}
}}
\caption{{Exploratory descriptor calculations on the saved species-mixture features (mean $\pm$ sample SD over three seeds). The original eight-descriptor score is unchanged. The six-descriptor score excludes volume and compactness with the same standardization and covariance conventions, and the closed subset keeps each sampler's own watertight outputs. Compare samplers within the original-eight or six-descriptor column; feature dimensions differ between columns. The closed-subset values are sensitivity checks and cannot rank samplers because selection and sample size differ.}}
\label{stab:tree_descriptor_sensitivity}
\end{table}

\paragraph{{Results.}} {The flow gives the lowest mean species-mixture MMD-CD over three seeds, 16.5\% below the per-species Gaussian and lower in all three seeds, and about 2\% below empirical residual sampling, which is lower in seed 43. Its nn-CD is the lowest in every seed. These mean differences were not tested for statistical significance.}

{The per-species Gaussian has lower MMD-CD than the flow for \textit{C.~parvulus}, \textit{G.~fuliginosa} and \textit{G.~magnirostris}, and the flow is lower for \textit{G.~fortis} (Table~\ref{stab:tree_focal}). Because the references are the training skulls, these results do not measure generalization to unseen specimens or species. Tests 1 and 3 of the generativity diagnostics in main-text Section~5.5 and Table~2 use the same 300 mixture outputs of the flow; Test 2 uses the proxy-based representation. These complementary diagnostics do not establish the absence of memorization.}

\paragraph{{Descriptor sensitivity.}} {Table~\ref{stab:tree_descriptor_sensitivity} recalculates descriptor scores from the saved feature arrays and leaves the meshes, models and locked MorphFID-8 unchanged. The six-descriptor version omits volume and compactness, which use the convex-hull fallback, and keeps area, the three bounding-box extents, elongation and flatness, which extraction geometry can still affect. The closed-subset version uses each sampler's own watertight mixture outputs against all 100 real references. Selection and sample size then differ by sampler and affect the finite-sample Fr\'echet estimates~\cite{chong2020effectively}, so this calculation tests sensitivity and cannot rank the samplers. In every run, the mean z-scored convex-hull volume of open outputs lies 1.5--2.5 reference SD above that of closed outputs from the same sampler, and compactness shows a similar shift. The flow has the most open outputs, which is consistent with its higher MorphFID-8.}

\section{{Proxy-Based Sampler Comparison}}
\label{sec:si_proxy}
{This comparison used the proxy-based representation and the Stage-2 schedule of Section~\ref{sec:si_impl}, with three flow seeds and one stored draw for each baseline. It includes a copying reference (Table~\ref{tab:historical_sampler}); Table~\ref{stab:paired} also compares the initial conditioned CFM.}

\begin{table}[t]
\centering\footnotesize
\resizebox{\ifdim\width>\linewidth\linewidth\else\width\fi}{!}{%
{\color{black}
\begin{tabular}{lcccccc}
\toprule
Sampler & nn-CD$\downarrow$ & MMD-CD$\downarrow$ & COV$\uparrow$ & 1-NNA$\to0.5$ & Diversity & MorphFID-8$\downarrow$\\
\midrule
Pooled residual flow & 1.37\,$\pm$\,0.03 & 1.25\,$\pm$\,0.03 & 0.50\,$\pm$\,0.02 & 0.59\,$\pm$\,0.04 & 10.59\,$\pm$\,0.12 & 1.13\,$\pm$\,0.24\\
Pooled Gaussian residual & 1.36 & 1.29 & 0.54 & 0.51 & 10.97 & 0.82\\
Empirical residual & 1.46 & 1.34 & 0.55 & 0.52 & 10.85 & 1.08\\
Per-species Gaussian (GMM) & 1.42 & 1.39 & 0.60 & 0.52 & 11.32 & 0.91\\
\midrule
Copying training skulls & 0.25 & 0.74 & 0.60 & 0.49 & 11.54 & n/a\\
Real skulls, leave-one-out & 1.47 & n/a & n/a & n/a & 11.48 & n/a\\
\bottomrule
\end{tabular}
}}
\caption{{Proxy-based species-mixture comparison (100 generated against 100 real skulls, Chamfer values $\times10^{-3}$). Flow values are mean $\pm$ SD over three seeds, and the other samplers use one stored draw. Copying and decoded references use the proxy-based decoder. The paired tests are in Table~\ref{stab:paired} and do not test the dated-tree results. n/a: not applicable.}}
\label{tab:historical_sampler}
\end{table}

\begin{table}[t]
\centering
\footnotesize
\resizebox{\ifdim\width>\linewidth\linewidth\else\width\fi}{!}{%
{\color{black}
\begin{tabular}{llcccc}
\toprule
Baseline & Metric & Seed 42 & Seed 43 & Seed 44 & {Seeds pooled}{}\\
\midrule
Pooled Gaussian residual & nn-CD$\downarrow$ & +0.036 (0.77) & -0.006 (0.96) & -0.014 (0.89) & {+0.005 (0.95)}{}\\
 & MMD-CD$\downarrow$ & -0.073 (0.60) & -0.006 (0.96) & -0.034 (0.77) & {-0.037 (0.56)}{}\\
 & COV$\uparrow$ & -0.030 (0.54) & -0.060 (0.19) & -0.025 (0.62) & {-0.038 (0.27)}{}\\
\addlinespace
Empirical residual & nn-CD$\downarrow$ & -0.065 (0.53) & -0.107 (0.35) & -0.114 (0.30) & {-0.095 (0.24)}{}\\
 & MMD-CD$\downarrow$ & -0.125 (0.10) & -0.058 (0.47) & -0.087 (0.24) & {-0.090 (0.09)}{}\\
 & COV$\uparrow$ & -0.045 (0.33) & -0.075 (0.11) & -0.040 (0.33) & {-0.053 (0.10)}{}\\
\addlinespace
Per-species Gaussian (GMM) & nn-CD$\downarrow$ & -0.019 (0.89) & -0.061 (0.64) & -0.068 (0.54) & {-0.049 (0.57)}{}\\
 & MMD-CD$\downarrow$ & -0.177 (0.02) & -0.110 (0.15) & -0.138 (0.06) & {-0.142 (0.010)}{}\\
 & COV$\uparrow$ & -0.095 (0.03) & -0.125 (0.005) & -0.090 (0.03) & {-0.103 (0.005)}{}\\
 & MorphFID-8$\downarrow$ & -0.04 (0.90) & +0.25 (0.47) & +0.44 (0.28) & {+0.22 (0.43)}{}\\
\addlinespace
Conditional CFM (initial design) & nn-CD$\downarrow$ & -0.682 ($<$0.001) & -0.724 (0.002) & -0.731 ($<$0.001) & {-0.712 ($<$0.001)}{}\\
 & MMD-CD$\downarrow$ & -0.528 ($<$0.001) & -0.461 ($<$0.001) & -0.490 ($<$0.001) & {-0.493 ($<$0.001)}{}\\
 & COV$\uparrow$ & +0.095 (0.02) & +0.065 (0.10) & +0.100 (0.01) & {+0.087 (0.008)}{}\\
 & {MorphFID-8}{}$\downarrow$ & {-1.74 ($<$0.001)}{} & {-1.44 (0.004)}{} & {-1.26 (0.02)}{} & {-1.48 ($<$0.001)}{}\\
\addlinespace
\bottomrule
\end{tabular}
}%
}
\caption{{Proxy-based paired permutation tests, pooled residual CFM minus baseline, per generator seed and pooled over seeds with swaps stratified by seed (species-mixture cohort, final decoder, 2,000 within-slot label swaps, two-sided $p$ with $+1$ correction, Chamfer differences $\times10^{-3}$). The MorphFID-8 row compares the per-species Gaussian, whose single sample set is shared across seeds. These are not tests of the dated-tree results.}}
\label{stab:paired}
\end{table}

\paragraph{{Results and tests.}} {The flow again has the lowest mixture MMD-CD (1.25, against 1.29, 1.34 and 1.39). The paired permutation tests of Table~\ref{stab:paired} swap the method labels of the two samples in each species-matched slot 2,000 times and give two-sided $p$-values with a $+1$ correction, with pooled swaps stratified by flow seed. Pooled over seeds, the flow's MMD-CD is lower than that of the per-species Gaussian ($-0.142$, $p=0.010$) but not detectably lower than that of empirical ($-0.090$, $p=0.09$) or pooled Gaussian ($-0.037$, $p=0.56$) residuals, and its COV is lower than the per-species Gaussian's ($-0.103$, $p=0.005$). Its nn-CD (1.37) is close to that of pooled Gaussian residuals (1.36). Against the initial conditioned CFM, the flow lowers nn-CD by 0.71 and MMD-CD by 0.49 (both $p<0.001$). The random-relabelling interval for 1-NNA in this comparison is 0.44--0.56, and the flow's 1-NNA ($0.59\pm0.04$) lies above it, whereas the three baselines lie inside it. These tests, references and intervals apply only to this comparison.}

\begin{table}[t]
\centering
\footnotesize
\resizebox{\ifdim\width>\linewidth\linewidth\else\width\fi}{!}{%
{\color{black}
\begin{tabular}{lcccccc}
\toprule
Sampler & nn-CD$\downarrow$ & MMD-CD$\downarrow$ & COV$\uparrow$ & Diversity & MorphFID-8$\downarrow$ & Within-species (\%)$\to100\%$\\
\midrule
Pooled residual CFM & 1.38\,$\pm$\,0.03 & 1.25\,$\pm$\,0.04 & 0.50\,$\pm$\,0.01 & 10.61\,$\pm$\,0.13 & 1.56\,$\pm$\,0.20 & 26/54/62/55\\
Pooled Gaussian residual & 1.37 & 1.29 & 0.54 & 10.99 & 1.06 & 59/54/75/44\\
Empirical residual & 1.47 & 1.35 & 0.54 & 10.86 & 1.48 & 39/59/80/43\\
Per-species Gaussian (GMM) & 1.42 & 1.40 & 0.60 & 11.31 & 1.22 & 80/93/107/73\\
Conditional CFM (initial design) & 2.07 & 1.74 & 0.41 & 11.45 & 2.78 & 19/21/55/25\\
\bottomrule
\end{tabular}
}%
}
\captionsetup{justification=raggedright,singlelinecheck=false}
\caption{{Proxy-based sampler comparison repeated with the lowest-training-loss decoder (Chamfer values $\times10^{-3}$; within-species order \textit{C.\,parvulus}/\allowbreak\textit{G.\,fortis}/\allowbreak\textit{G.\,fuliginosa}/\allowbreak\textit{G.\,magnirostris}).}}
\label{stab:best}
\end{table}

\begin{table}[t]
\centering
\footnotesize
\resizebox{\ifdim\width>\linewidth\linewidth\else\width\fi}{!}{%
{\color{black}
\begin{tabular}{lcccccc}
\toprule
& \multicolumn{4}{c}{MorphFID$\downarrow$} & \multicolumn{2}{c}{Topology (\%)}\\
\cmidrule(lr){2-5}\cmidrule(lr){6-7}
Sampler & 8 features & 11 features & 12 features & 12, physical frame & open$\downarrow$ & multi-part\\
\midrule
Pooled residual CFM & 1.13\,$\pm$\,0.24 & 1.46\,$\pm$\,0.24 & 14556 & 14143 & 33 & 45\\
Pooled Gaussian residual & 0.82 & 0.99 & 14081 & 13685 & 30 & 47\\
Empirical residual & 1.08 & 1.27 & 14070 & 13674 & 35 & 46\\
Per-species Gaussian (GMM) & 0.91 & 1.36 & 13451 & 13077 & 20 & 31\\
Conditional CFM (initial design) & 2.61 & 6.09 & 11773 & 11455 & 24 & 45\\
{Decoded training codes (reference)}{} & {n/a}{} & {n/a}{} & {n/a}{} & {n/a}{} & {11}{} & {28}{}\\
\bottomrule
\end{tabular}
}%
}
\caption{{Proxy-based MorphFID variants and mesh topology for the species-mixture cohort (final decoder). The eight geometric descriptors are the primary set. The 11-descriptor set adds the three centroid coordinates, and the 12-descriptor set also adds the vertex count, which depends on grid resolution and dominates the score. The physical-frame variant compares raw specimens in millimetres with normalized generated meshes. Open: the surface meets the boundary of $[-1,1]^3$. Multi-part: more than one connected component, without filtering. The last row gives the topology of the 100 decoded training skulls.}}
\label{stab:mfid}
\end{table}

\paragraph{{Lowest-training-loss decoder.}} {Repeating the comparison with the checkpoint of lowest training loss gives the same ordering of mixture MMD-CD (Table~\ref{stab:best}).}

\begin{figure}[t]
\centering
\includegraphics[width=\linewidth]{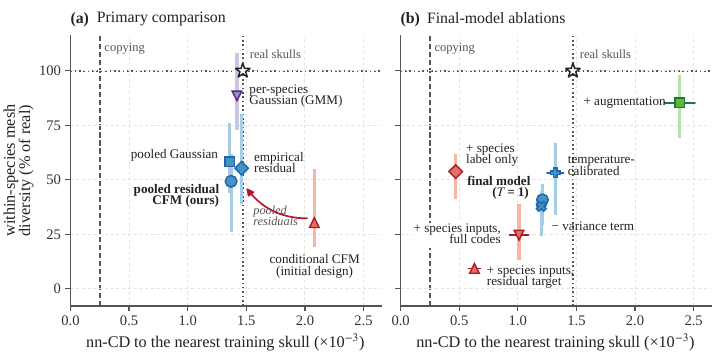}
\caption{{
\textbf{Proxy-based fidelity and within-species mesh diversity.}
(a)~Sampler comparison (Table~\ref{tab:historical_sampler},
with diversity in Table~\ref{stab:diversity}).
(b)~Final-model ablations (Table~\ref{stab:s03}).
The vertical coordinate is the mean diversity percentage over
the four focal species, vertical bars span the minimum and
maximum, and horizontal bars show the seed SD of repeated
flow runs. Other samplers in panel~(a) use one stored draw.
The panels use different training and sampling schedules,
so comparisons are made within each panel.
}}
\label{fig:historical_fidelity}
\end{figure}

\paragraph{{Diversity and decoder diagnostics.}} {Table~\ref{stab:diversity} gives within-species mesh diversity. Decoding the training codes keeps 98--100\% of real within-species pairwise diversity for the four focal species. At $T=1$, the flow's sampled residual variance is 0.48--0.57 of the training residual variance across three seeds, against about 1.0 for the pooled Gaussian and empirical samplers and 2.0 for the per-species Gaussian. These values implicate the generator in the reduced variation under this protocol. Reconstruction at fitted codes cannot rule out decoder compression along other latent directions, and the check was not repeated for the tree-trained decoder. Figure~\ref{fig:historical_fidelity} plots fidelity against diversity for this comparison and for the final-model ablations.}

\begin{table}[t]
\centering
\footnotesize
{\color{black}
\begin{tabular}{lcccc}
\toprule
Diversity (\%)$\to100\%$ & \textit{C.\,parvulus} & \textit{G.\,fortis} & \textit{G.\,fuliginosa} & \textit{G.\,magnirostris}\\
\midrule
Real skulls, pairwise CD ($\times10^{-3}$) & 9.74 & 7.35 & 7.04 & 6.13\\
Decoded real skulls (\%) & 100 & 98 & 99 & 99\\
\midrule
\textbf{Pooled residual CFM (ours)} & 26 & 54 & 62 & 55\\
Pooled Gaussian residual & 59 & 54 & 76 & 44\\
Empirical residual & 39 & 59 & 80 & 43\\
Per-species Gaussian (GMM) & 79 & 94 & 108 & 73\\
Conditional CFM (initial design) & 19 & 21 & 55 & 26\\
Ours, temperature-calibrated$^{\dagger}$ & 34 & 54 & 58 & 67\\
\bottomrule
\end{tabular}
}
\caption{{Proxy-based within-species mesh diversity as a percentage of the real species' pairwise CD (first row), for ten generated skulls per species. Decoding the real training codes keeps 98--100\% of the measured real diversity at those codes, which does not exclude decoder effects along generated latent directions. $^{\dagger}$Ablation schedule (main-text Section~5.6 and Fig.~7b), under which the uncalibrated flow gives 29/40/48/46\%.}}
\label{stab:diversity}
\end{table}

\section{{Component Studies}}
\label{sec:si_components}
\paragraph{{Initial design.}} {Table~\ref{stab:s02} retrains the network of the initial design for 10,000 updates (batch 256, seed 42) on the 1,107 augmented codes described below, with original codes given three-fold weight in training draws and guidance scale 1.5 at generation. It also reports MorphFID-8, per-species diversity and simple samplers evaluated under the same protocol.}

\begin{table}[t!]
\centering
\footnotesize
\setlength{\tabcolsep}{2.5pt}
\renewcommand{\arraystretch}{1.18}
\par\smallskip\noindent\textbf{(a) Generator performance}\par\smallskip
\begin{tabular*}{\linewidth}{@{\extracolsep{\fill}}lrrrrrrrr@{}}
\toprule
 & \multicolumn{4}{c}{Species-mixture metrics} & \multicolumn{4}{c}{Within-species (\%)}\\
\cmidrule(lr){2-5}\cmidrule(l){6-9}
Configuration & nn-CD$\downarrow$ & MMD-CD$\downarrow$ & COV$\uparrow$ & MorphFID-8$\downarrow$ & \textit{C.p.} & \textit{G.fo.} & \textit{G.fu.} & \textit{G.m.}\\
\midrule
Initial design & 2.07 & 1.59 & 0.560 & 2.00 & 54 & 79 & 69 & 86\\
\addlinespace[3pt]
No phylogenetic conditioning & 2.05 & 1.65 & 0.550 & 2.04 & 59 & 87 & 69 & 84\\
No variance term & 1.81 & 1.55 & 0.540 & 1.47 & 47 & 83 & 63 & 75\\
No residual target & 2.01 & 1.54 & 0.480 & 3.19 & 45 & 45 & 74 & 59\\
Standard CFM & 1.57 & 1.45 & 0.505 & 1.98 & 42 & 44 & 66 & 53\\
No augmentation & 0.53 & 1.25 & 0.380 & 0.59 & 14 & 7 & 9 & 6\\
\addlinespace[3pt]
Empirical pooled residual & 1.31 & 1.20 & 0.565 & 0.68 & 48 & 63 & 76 & 46\\
Per-species shrinkage Gaussian & 1.41 & 1.29 & 0.610 & 0.89 & 66 & 114 & 66 & 83\\
Pooled full-code Gaussian$^{a}$ & 1.68 & 1.68 & 0.500 & 0.71 & 67 & 138 & 105 & 140\\
EM diagonal GMM$^{b}$ & 1.58 & 1.79 & 0.425 & 1.14 & \multicolumn{4}{c}{n/a}\\

\bottomrule
\end{tabular*}
\medskip
\par\smallskip\noindent\textbf{(b) Paired comparisons with the initial design}\par\smallskip
\begin{tabular*}{\linewidth}{@{\extracolsep{\fill}}lrrrr@{}}
\toprule
 & \multicolumn{2}{c}{nn-CD} & \multicolumn{2}{c}{MMD-CD}\\
\cmidrule(lr){2-3}\cmidrule(l){4-5}
Configuration & $\Delta\downarrow$ & $p$ & $\Delta\downarrow$ & $p$\\
\midrule
Initial design & n/a & n/a & n/a & n/a\\
\addlinespace[3pt]
No phylogenetic conditioning & -0.02 & 0.56 & +0.06 & 0.32\\
No variance term & -0.26 & $<$0.001 & -0.04 & 0.28\\
No residual target & -0.06 & 0.76 & -0.05 & 0.49\\
Standard CFM & -0.50 & 0.0020 & -0.14 & 0.02\\
No augmentation & -1.55 & $<$0.001 & -0.34 & $<$0.001\\
\addlinespace[3pt]
Empirical pooled residual & -0.76 & $<$0.001 & -0.40 & $<$0.001\\
Per-species shrinkage Gaussian & -0.66 & $<$0.001 & -0.30 & $<$0.001\\
Pooled full-code Gaussian$^{a}$ & -0.39 & 0.03 & +0.09 & 0.30\\
EM diagonal GMM$^{b}$ & -0.49 & 0.0015 & +0.20 & 0.05\\

\bottomrule
\end{tabular*}
\caption{{Proxy-based matched generator study: the network of the initial design (Conditional CFM) retrained for 10,000 updates (batch 256, seed 42), with 140 attempts per arm (100 for the EM Gaussian mixture). In panel~(b), differences are arm minus the initial design (Chamfer values $\times10^{-3}$), with paired permutation tests on the species-mixture cohort. The initial design is the reference row (n/a, not applicable). Within-species diversity has a target of 100\%. The focal-species columns are \textit{C.~parvulus} (C.p.), \textit{G.~fortis} (G.fo.), \textit{G.~fuliginosa} (G.fu.) and \textit{G.~magnirostris} (G.m.). $^{a}$Negative control. $^{b}$Mixture only.}  }
\label{stab:s02}
\end{table}

\begin{table}[t!]
\centering
\footnotesize
\setlength{\tabcolsep}{2.5pt}
\renewcommand{\arraystretch}{1.18}
\par\smallskip\noindent\textbf{(a) Species-mixture metrics: mean $\pm$ SD}\par\smallskip
\begin{tabular*}{\linewidth}{@{\extracolsep{\fill}}lrrrr@{}}
\toprule
Arm & nn-CD$\downarrow$ & MMD-CD$\downarrow$ & COV$\uparrow$ & MorphFID-8$\downarrow$\\
\midrule
Final, $T=1$ & 1.21\,$\pm$\,0.04 & 1.24\,$\pm$\,0.06 & 0.49\,$\pm$\,0.02 & 0.86\,$\pm$\,0.06\\
Temperature-calibrated & 1.32\,$\pm$\,0.07 & 1.25\,$\pm$\,0.02 & 0.55\,$\pm$\,0.02 & 0.91\,$\pm$\,0.22\\
No variance term & 1.20\,$\pm$\,0.04 & 1.24\,$\pm$\,0.05 & 0.49\,$\pm$\,0.02 & 1.00\,$\pm$\,0.21\\
With augmentation & 2.38\,$\pm$\,0.13 & 1.77\,$\pm$\,0.08 & 0.52\,$\pm$\,0.02 & 3.61\,$\pm$\,0.64\\
\addlinespace[4pt]
\multicolumn{5}{l}{\textit{Species-conditioned variants (corrected size inputs where used)}}\\
Species inputs: residual & 0.63\,$\pm$\,0.05 & 1.15\,$\pm$\,0.03 & 0.42\,$\pm$\,0.03 & 0.68\,$\pm$\,0.05\\
Species inputs: full codes & 1.01\,$\pm$\,0.08 & 1.12\,$\pm$\,0.05 & 0.51\,$\pm$\,0.01 & 1.11\,$\pm$\,0.25\\
Species label only & 0.47\,$\pm$\,0.04 & 0.71\,$\pm$\,0.01 & 0.64\,$\pm$\,0.02 & 0.91\,$\pm$\,0.30\\
\addlinespace[4pt]
\multicolumn{5}{l}{\textit{Superseded runs: misaligned size inputs}}\\
Species inputs: residual & 0.70\,$\pm$\,0.05 & 1.15\,$\pm$\,0.09 & 0.43\,$\pm$\,0.03 & 0.78\,$\pm$\,0.18\\
Species inputs: full codes & 0.85\,$\pm$\,0.02 & 1.13\,$\pm$\,0.04 & 0.48\,$\pm$\,0.04 & 0.86\,$\pm$\,0.17\\

\bottomrule
\end{tabular*}
\medskip
\par\smallskip\noindent\textbf{(b) Within-species diversity (\%; target 100)}\par\smallskip
\begin{tabular*}{\linewidth}{@{\extracolsep{\fill}}lrrrr@{}}
\toprule
Arm & \textit{C.~parvulus} & \textit{G.~fortis} & \textit{G.~fuliginosa} & \textit{G.~magnirostris}\\
\midrule
Final, $T=1$ & 29 & 40 & 48 & 46\\
Temperature-calibrated & 34 & 54 & 58 & 67\\
No variance term & 24 & 39 & 45 & 43\\
With augmentation & 69 & 83 & 91 & 98\\
\addlinespace[4pt]
\multicolumn{5}{l}{\textit{Species-conditioned variants (corrected size inputs where used)}}\\
Species inputs: residual & 8 & 10 & 8 & 11\\
Species inputs: full codes & 13 & 21 & 39 & 25\\
Species label only & 41 & 62 & 61 & 51\\
\addlinespace[4pt]
\multicolumn{5}{l}{\textit{Superseded runs: misaligned size inputs}}\\
Species inputs: residual & 23 & 5 & 10 & 15\\
Species inputs: full codes & 28 & 16 & 33 & 29\\

\bottomrule
\end{tabular*}
\caption{{Proxy-based final-model ablation and corrected-size runs, mean $\pm$ SD over generator seeds 42--44 on one representation (Chamfer values $\times10^{-3}$, within-species order as in Table~\ref{stab:best}). Rows marked superseded used misaligned size inputs.}}
\label{stab:s03}
\end{table}

\paragraph{{Final-model ablations.}} {Table~\ref{stab:s03} gives the three-seed variants of the final model, summarized in main-text Section~5.6, with MorphFID-8 and per-species diversity. The initial design read size inputs that were misaligned for 88 skulls. The conditioned variants were rerun with corrected sizes, and the superseded rows are listed for comparison. The final model fixes the size input at zero and is unaffected. Standard CFM in the matched initial-design study retains 42--66\% of real focal-species diversity (Table~\ref{stab:s02})}.

\paragraph{{Latent augmentation as run.}} {The initial design expanded the 100 codes to 1,107. It added 500 perturbed codes, five per skull, each formed by adding Gaussian noise with SD 0.1 per dimension, reweighted across dimensions by the inverse of a finite-difference estimate of the diagonal of the decoder Hessian, with principal components used only as a fallback. It also added 435 interpolations between species pairs with proxy distance at most 0.7 and 72 extrapolations of species centroids away from the global mean ($\gamma\in\{0.3,0.5,0.7\}$), clipped to the range of the real codes $\pm3$ SD. Before reweighting, the isotropic noise has root-mean-square norm $0.1\sqrt{64}=0.8$, about twice the mean residual norm of 0.38 and close to the mean distance between centroids (0.90). The final model uses no augmentation.}

\paragraph{{Initial-design sampler.}} {The initial conditional flow used guidance scale 1.5 and 50 Euler steps. The effective temperature was 1, and the original generation code left the size input at zero.}

\paragraph{{Diffusion baseline.}} {In the initial DDPM experiment with 1,000 denoising steps, the standard deviation of generated codes ranged from $0.17\times$ to $338\times$ the reference value across epochs, and no valid 3D meshes were extracted. Its inputs and configuration could not be recovered, so it was not rerun under the matched protocol. This run does not establish a general limitation of diffusion models.}

\section{{Phylogenetic Supervision Controls}}
\label{sec:si_supervision}
{For each supervising matrix (Section~\ref{sec:si_matrices}), the representation was retrained for 2,000 epochs with seed 42, identical initial weights and the same data order. A pooled flow was then trained on each arm's codes with the ablation schedule and seed 42, and 100 mixture skulls and ten skulls for each focal species were generated and evaluated.    Table~\ref{stab:phylo_controls} gives LPC, centroid correlations and mixture results, and Table~\ref{stab:focal_controls} gives the focal-species results. Only the proxy arm differs detectably from the no-loss arm in focal MMD-CD (14 of 21 real skulls closer, exact paired sign-flip $p=0.008$). The proxy arm also differs from shuffled proxy 3 ($p=0.012$) but not from shuffled proxies 1 and 2 ($p=0.18$) or the dated tree ($p=0.13$). With one representation seed, these controls show no generation benefit specific to the correct tree.}

\begin{table}[H]
\centering
\footnotesize
\resizebox{\ifdim\width>\linewidth\linewidth\else\width\fi}{!}{%
{\color{black}
\begin{tabular}{lccccccc}
\toprule
\textbf{Supervising matrix} & \textbf{LPC}$\uparrow$ & \textbf{$r\uparrow$, proxy ($p$)} & \textbf{$r\uparrow$, tree ($p$)} & \textbf{nn-CD$\downarrow$} & \textbf{MMD-CD$\downarrow$} & {\textbf{COV$\uparrow$}}{} & {\textbf{Diversity}}{}\\
\midrule
None ($\beta=0$) & 0.23$^{\ddagger}$ & 0.26 ($<$0.001) & 0.17 (0.12) & 1.27 & 1.18 & {0.56}{} & {10.37}{}\\
Taxonomic proxy & 0.99 & 0.99 ($<$0.001) & 0.29 ($<$0.001) & 1.20 & 1.22 & {0.52}{} & {10.05}{}\\
Dated tree & 0.99 & 0.35 ($<$0.001) & 0.99 ($<$0.001) & 1.22 & 1.29 & {0.46}{} & {9.93}{}\\
Shuffled proxy 1 & 0.99 & -0.02 (0.54) & -0.04 (0.80) & 1.13 & 1.16 & {0.52}{} & {10.08}{}\\
Shuffled proxy 2 & 0.99 & -0.04 (0.69) & 0.02 (0.30) & 1.19 & 1.21 & {0.49}{} & {9.75}{}\\
Shuffled proxy 3 & 0.99 & -0.01 (0.44) & -0.04 (0.79) & 1.24 & 1.24 & {0.47}{} & {10.00}{}\\
\bottomrule
\end{tabular}
}%
}
\caption{{\textbf{Phylogenetic supervision controls.} DeepSDF retrained for 2,000 epochs (seed~42) under each supervising matrix. LPC: specimen-level correlation between latent and supervising distances. $r$: correlation of the 276 species-centroid distances with the proxy or tree, with one-sided $p$ from 9,999 label permutations. Generation: 100 mixture samples from a Residual CFM retrained on each arm's codes (Chamfer values $\times10^{-3}$). Shuffled matrices are derangements of the proxy with $|r|<0.1$ to both reference matrices. $^{\ddagger}$Correlation with the proxy.}}
\label{stab:phylo_controls}
\end{table}

\begin{table}[H]
\centering
\footnotesize
{\color{black}
\begin{tabular}{lccccccc}
\toprule
Supervising matrix & nn-CD$\downarrow$ & MMD-CD$\downarrow$ & COV$\uparrow$ & Diversity & MorphFID-8$\downarrow$ & Closer$\uparrow$ & $p$\\
\midrule
None ($\beta=0$) & 1.75 & 1.81 & 0.55 & 2.89 & 4.51 & n/a & n/a\\
Taxonomic proxy & 1.58 & 1.47 & 0.66 & 3.16 & 5.43 & 14/21 & 0.008\\
Dated tree & 1.81 & 1.77 & 0.58 & 2.77 & 4.06 & 10/21 & 0.70\\
Shuffled proxy 1 & 1.42 & 1.73 & 0.55 & 2.67 & 4.05 & 12/21 & 0.48\\
Shuffled proxy 2 & 1.68 & 1.71 & 0.62 & 2.66 & 4.40 & 12/21 & 0.21\\
Shuffled proxy 3 & 1.99 & 1.98 & 0.48 & 2.34 & 3.97 & 12/21 & 0.55\\
\bottomrule
\end{tabular}
}
\caption{{Supervision controls of Table~\ref{stab:phylo_controls} for the four focal species: ten generated skulls per species and arm, species means (Chamfer values $\times10^{-3}$; diversity is the mean pairwise distance among a species' generated skulls; MorphFID-8 on the eight geometric descriptors, lower is better). Closer and $p$ compare, for each of the 21 real focal skulls, the distance to its nearest generated skull with the no-loss value, giving the number of skulls with a closer generated skull and the two-sided exact paired sign-flip $p$. One representation seed. n/a: not applicable.}}
\label{stab:focal_controls}
\end{table}

\section{{Withheld Species}}
\label{sec:si_heldout}
{The representation was retrained on 79 skulls of 20 species, withholding all 21 skulls of \textit{C.~parvulus}, \textit{G.~fortis}, \textit{G.~fuliginosa} and \textit{G.~magnirostris}. We distinguish prediction without withheld scans, reconstruction with codes fitted to those scans, and oracle generation using their fitted centroids.}

\paragraph{{Held-out generation.}} {A centroid is predicted from training centroids alone. For a supplied distance matrix $D$, let $\mathcal{N}_3(s)$ contain the three nearest training species, with ties broken alphabetically. The predictor is
\[
\widehat{\boldsymbol{\mu}}_s =
\frac{\sum_{j\in\mathcal{N}_3(s)}w_{sj}\boldsymbol{\mu}_j}
     {\sum_{j\in\mathcal{N}_3(s)}w_{sj}},
\qquad w_{sj}=\frac{1}{D_{sj}+10^{-6}},
\]
applied to the stored tree matrix, the proxy and, where applicable, the shuffled training matrix. The global-mean control averages all training-species centroids equally. The pooled flow is trained only on the 79 training codes. No withheld scans enter centroid prediction or residual generation; they are used only for evaluation. The same sampled residuals are added to each predicted centroid to give ten skulls per withheld species, evaluated against that species' real skulls.}

\paragraph{{Oracle generation.}} {The oracle centroid is the mean of the fitted codes of the withheld species. It uses the test geometry, and its generated skulls are evaluated against the same scans, so it measures consistency with observed specimens rather than prediction.}

{Table~\ref{stab:heldout} covers all predictors, Table~\ref{stab:knn_paired} compares tree and proxy neighbours sample by sample, and Table~\ref{stab:heldout_main} lists the values plotted in main-text Fig.~8. The experiment uses one split, one representation seed and four withheld species.}

\begin{table}[t]
\centering
\scriptsize
\resizebox{\ifdim\width>\linewidth\linewidth\else\width\fi}{!}{%
{\color{black}
\begin{tabular}{llccc}
\toprule
Trained with & Centroid predictor & nn-CD$\downarrow$ & MMD-CD$\downarrow$ & COV$\uparrow$\\
\midrule
None ($\beta=0$) & global mean & 4.04 & 4.84 & 0.60\\
 & proxy $k$NN & 4.76 & 6.33 & 0.46\\
 & tree $k$NN & 3.75 & 4.94 & 0.52\\
 & oracle & 2.02 & 1.97 & 0.52\\
\addlinespace
Taxonomic proxy & global mean & 3.89 & 4.79 & 0.48\\
 & proxy $k$NN & 4.40 & 6.50 & 0.39\\
 & tree $k$NN & 3.53 & 5.26 & 0.45\\
 & oracle & 2.40 & 2.29 & 0.48\\
\addlinespace
Dated tree & global mean & 3.49 & 4.20 & 0.43\\
 & proxy $k$NN & 4.09 & 5.20 & 0.41\\
 & tree $k$NN & 3.26 & 4.17 & 0.48\\
 & oracle & 2.42 & 2.21 & 0.48\\
\addlinespace
Shuffled tree 1 & global mean & 4.29 & 5.28 & 0.51\\
 & proxy $k$NN & 4.25 & 6.25 & 0.46\\
 & tree $k$NN & 3.46 & 5.19 & 0.44\\
 & own-matrix $k$NN & 4.79 & 6.49 & 0.53\\
 & oracle & 3.00 & 2.18 & 0.53\\
\addlinespace
Shuffled tree 2 & global mean & 4.06 & 4.44 & 0.46\\
 & proxy $k$NN & 4.20 & 5.28 & 0.43\\
 & tree $k$NN & 3.48 & 4.41 & 0.52\\
 & own-matrix $k$NN & 5.75 & 6.99 & 0.41\\
 & oracle & 2.32 & 2.32 & 0.48\\
\addlinespace
Shuffled tree 3 & global mean & 3.81 & 4.70 & 0.51\\
 & proxy $k$NN & 4.21 & 6.31 & 0.43\\
 & tree $k$NN & 3.37 & 4.93 & 0.52\\
 & own-matrix $k$NN & 5.45 & 6.43 & 0.48\\
 & oracle & 2.45 & 2.32 & 0.62\\
\addlinespace
\bottomrule
\end{tabular}
}%
}
\caption{{Withheld-species generation for all centroid predictors (means over four species, ten samples each, Chamfer values $\times10^{-3}$).}}
\label{stab:heldout}
\end{table}

\begin{table}[t]
\centering
\footnotesize
{\color{black}
\begin{tabular}{lcccc}
\toprule
Trained with & nn-CD$\uparrow$ & MMD-CD$\uparrow$ & \textit{C.~parvulus} nn-CD$\uparrow$ & \textit{Geospiza} nn-CD$\uparrow$\\
\midrule
None ($\beta=0$) & +1.01 ($<$0.001) & +1.39 (0.003) & +3.09 & +0.31 (0.07)\\
Taxonomic proxy & +0.87 ($<$0.001) & +1.24 (0.01) & +3.07 & +0.14 (0.26)\\
Dated tree & +0.82 ($<$0.001) & +1.02 (0.001) & +2.50 & +0.27 (0.04)\\
Shuffled tree 1 & +0.80 ($<$0.001) & +1.06 (0.03) & +2.92 & +0.09 (0.32)\\
Shuffled tree 2 & +0.72 ($<$0.001) & +0.87 (0.01) & +2.17 & +0.24 (0.04)\\
Shuffled tree 3 & +0.85 ($<$0.001) & +1.38 (0.002) & +3.12 & +0.09 (0.24)\\
\bottomrule
\end{tabular}
}
\caption{{Withheld species: proxy-neighbour minus tree-neighbour centroid prediction (positive values mean that tree neighbours are closer to the real withheld skulls; Chamfer values $\times10^{-3}$, species means as in Table~\ref{stab:heldout}). Within each representation the same 40 sampled residuals are added to every predicted centroid, so nn-CD is compared by generated skull (40 pairs, 400,000 random sign flips) and MMD-CD by real withheld skull (21 pairs, exact sign-flip test), both two-sided. The proxy neighbours of \textit{C.~parvulus} include \textit{Coereba flaveola}, read as \textit{Camarhynchus} (Section~\ref{sec:si_matrices}). The three withheld \textit{Geospiza} receive identical proxy predictions because of tied proxy distances, and the last column excludes \textit{C.~parvulus}. In the three representations not trained on a shuffled matrix, proxy neighbours were also worse than the global mean (nn-CD $+0.51$ to $+0.72$, $p\le0.03$).}}
\label{stab:knn_paired}
\end{table}


\section{{Morphometric Audit}}
\label{sec:si_anat}
\begin{figure}[t]
    \centering
    \includegraphics[width=\linewidth]{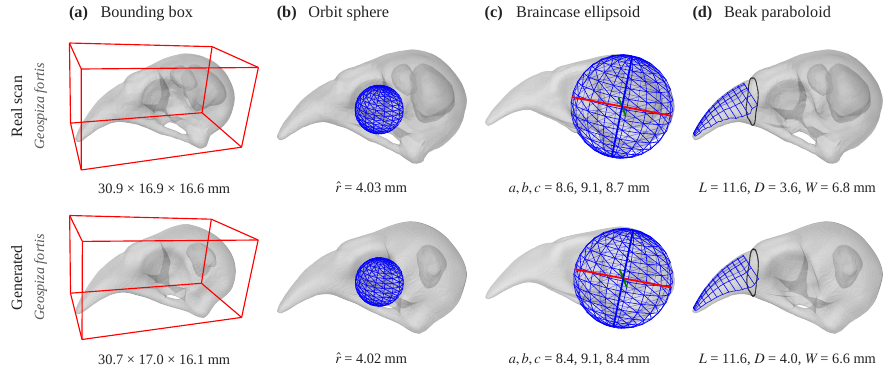}
    \caption{{\textbf{The automated skull measurements}, drawn with the figure code of~\cite{lau2025robust} on a raw scan (top) and on a generated \textit{Geospiza fortis} skull from the proxy-based morphometric audit (bottom; pooled residual flow, $T=1$, seed 42). This audit uses a different representation and sampling protocol from main-text Fig.~5. (a)~Axis-aligned bounding box (skull length, width and height). (b)~Sphere fitted to the orbit. (c)~Axis-aligned ellipsoid fitted to the braincase, with semi-axes $a$ (red, along the length), $b$ (green, transverse) and $c$ (blue, vertical). (d)~Upper beak surface fitted to the distal 30\% of skull length with the paraboloid model of~\cite{al2021geometry,mosleh2023beak}, with the cross-section at that plane in black. Values are the audit records in mm, each column is drawn at one scale, and for the scan the beak fit reproduces the recorded values.}}
    \label{sfig:measures}
\end{figure}

{We measured every skull with the code of~\cite{lau2025robust} (\url{https://github.com/kaikwanlau/skull-morphology}): skull length, width and height, orbit radius and curvature from a fitted sphere, and three braincase axes from a fitted ellipsoid. The upper beak was measured by fitting the paraboloid model of~\cite{al2021geometry,mosleh2023beak} to the distal 30\% of skull length. Figure~\ref{sfig:measures} shows the measurements on a raw scan and a generated skull. The 12 size-free quantities are skull width and height, orbit radius and the three braincase axes divided by skull length, beak depth and width divided by beak length, the dimensionless beak curvature and shape parameters, beak length divided by skull length, and the beak-axis angle.}

\begin{table}[t!]
\centering
\footnotesize
\resizebox{\ifdim\width>\linewidth\linewidth\else\width\fi}{!}{%
{\color{black}
\begin{tabular}{lccccc}
\toprule
& & \multicolumn{4}{c}{\textbf{Generation: nn-CD$\downarrow$\,/\,MMD-CD$\downarrow$}}\\
\cmidrule(lr){3-6}
\textbf{Trained with} & \textbf{Fitted recon.$\downarrow$\ ($p$)} & \textbf{Global mean} & \textbf{Tree $k$NN} & \textbf{Own-matrix $k$NN} & \textbf{Oracle}\\
\midrule
None ($\beta=0$) & 0.525 & 4.04\,/\,4.84 & 3.75\,/\,4.94 & n/a & 2.02\,/\,1.97\\
Taxonomic proxy & 0.495 (0.06) & 3.89\,/\,4.79 & 3.53\,/\,5.26 & 4.40\,/\,6.50 & 2.40\,/\,2.29\\
Dated tree & 0.487 (0.0045) & 3.49\,/\,4.20 & 3.26\,/\,4.17 & $=$ tree $k$NN & 2.42\,/\,2.21\\
Shuffled tree 1 & 0.487 (0.04) & 4.29\,/\,5.28 & 3.46\,/\,5.19 & 4.79\,/\,6.49 & 3.00\,/\,2.18\\
Shuffled tree 2 & 0.502 (0.12) & 4.06\,/\,4.44 & 3.48\,/\,4.41 & 5.75\,/\,6.99 & 2.32\,/\,2.32\\
Shuffled tree 3 & 0.495 (0.17) & 3.81\,/\,4.70 & 3.37\,/\,4.93 & 5.45\,/\,6.43 & 2.45\,/\,2.32\\
\bottomrule
\end{tabular}
}%
}
\caption{{Withheld-species values plotted in main-text Fig.~8. DeepSDF retrained on 79 skulls of 20 species (seed 42), with the 21 skulls of four species withheld. Fitted reconstruction: mean Chamfer distance after fitting codes to the withheld skulls with the decoder frozen, with an exact paired sign-flip test against $\beta=0$. Generation: species means over the four withheld species, ten samples each (Chamfer values $\times10^{-3}$), from centroids predicted from the three nearest training species in the dated tree or in the training matrix, or from the global mean. The oracle uses centroids fitted to the withheld scans and is a diagnostic using observed geometry, rather than a prediction without scans. The no-loss representation has no training matrix (n/a, not applicable).}}
\label{stab:heldout_main}
\end{table}

{The audit uses proxy-based outputs from the ablation schedule and decoded real controls. Real and generated codes are decoded on a common grid with 39 cells along skull length (the published remeshing parameter of 60 scaled by 0.65 to match the vertex counts of the published meshes), over a domain slightly larger than $[-1,1]^3$ so that beak tips are not cut. Meshes are converted to millimetres with each real skull's own scale, or with the species' mean scale for generated skulls. The pipeline reproduces the released measurements of the 100 real skulls to within $1.5\times10^{-6}$\,mm.}

\paragraph{{Results.}} {Of the 420 final-model skulls, 419 have all 12 measurements (Table~\ref{stab:anat}), and 98.5\% of their 5,028 values lie within two pooled reference SD. The median correlation of species means across the 12 measures is 0.98 (minimum 0.65). Correlations between generated and decoded-real species means are $r=0.982$ for beak depth/beak length, $r=0.996$ for beak width/beak length and $r=0.993$ for beak length/skull length (main-text Fig.~11b), showing preservation of these measured species-level beak proportions. The median orbit-rule error is 5.8\%. Pooled within-species SD is 55\% of the decoded-real reference. Temperature calibration raises it to 73\%, with a median orbit-rule error of 6.5\%. These automated geometric measurements have not been supplemented by expert anatomical assessment.}

\paragraph{{Displayed examples.}} {The generated \textit{C.~parvulus} skull of main-text Figs.~9 and~10 comes from the final pooled model trained with the ablation schedule ($T=1$, seed 42, slot 000). The orbit sphere (radius 3.18\,mm) and braincase ellipsoid are archived audit fits drawn with the code of~\cite{lau2025robust}. The displayed mesh was re-extracted on the audit grid from the same sample and differs from the archived audit mesh at 4 of 3,452 vertices and by 0.13\% in extent. The local beak axis joins the tip to the mean position of vertices in a slab about the distal-30\% cut. Upper-surface selection requires vertices above the fitted tomium with dorsally directed normals, and the lateral coordinate is detrended before the paraboloid fit. Depth is the fitted upper centreline minus the tomium at the base, and width averages the final three valid axial slices. The unchanged archived routine gives $L=9.20$\,mm, $D=2.92$\,mm, $W=6.45$\,mm and 242 selected vertices on the displayed mesh, against $L=9.19$\,mm, $D=2.92$\,mm, $W=6.45$\,mm and 237 vertices in the archived record. The 30\% cut is a geometric convention rather than a manually placed landmark.}

\begin{table}[t]
\centering
\footnotesize
\setlength{\tabcolsep}{2.5pt}
\renewcommand{\arraystretch}{1.18}
\begin{tabular*}{\linewidth}{@{\extracolsep{\fill}}lrrrrrrrr@{}}
\toprule
 & & \multicolumn{3}{c}{Fit and threshold percentages} & Curvature & \multicolumn{2}{c}{Species-mean $r$} & Within-sp.\\
\cmidrule(lr){3-5}\cmidrule(lr){7-8}
Skull set & $n$ & \shortstack{Orbit fit\\$\uparrow$} & \shortstack{Within\\$\pm2$ SD} & \shortstack{Below nominal\\$\chi^2_{12,0.95}$} & \shortstack{Median error\\(\%) $\downarrow$} & Median$\uparrow$ & Min.$\uparrow$ & \shortstack{SD (\% of\\decoded)}\\
\midrule
Raw scans & 100 & 100.0 & 90.5 & 77.8 & 5.3 & 0.94 & 0.52 & 101\\
Decoded real (reference) & 100 & 100.0 & 97.2 & 96.0 & 8.2 & \multicolumn{2}{c}{reference} & 100\\
\addlinespace[4pt]
Final model, $T=1$ & 420 & 99.8 & 98.5 & 95.2 & 5.8 & 0.98 & 0.65 & 55\\
Temperature-calibrated & 420 & 99.5 & 97.8 & 91.9 & 6.5 & 0.97 & 0.69 & 73\\
With augmentation & 420 & 92.6 & 92.4 & 58.1 & 12.1 & 0.94 & 0.46 & 114\\

\bottomrule
\end{tabular*}
\caption{{Proxy-based morphometric audit on 12 size-free measures, distinct from the dated-tree sampler comparison. The reference is the decoded real skulls (species means and pooled within-species SD and covariance). The column $n$ counts all skulls. The two threshold percentages use complete cases with all 12 measurements: 99/100 raw scans, 100/100 decoded real skulls, 419/420 final-model skulls, 418/420 temperature-calibrated skulls and 389/420 augmented skulls. One raw scan lacks a beak fit, and the generated exclusions lack an orbit radius. Orbit-fit percentages use all $n$ skulls, and curvature errors use available finite values. Generated arms pool three training seeds of the ablation schedule (140 skulls each). Curvature rule: orbit curvature predicted from skull length, width and height by Eq.~(8) of~\cite{lau2025robust}. Species-mean $r$: for each measure, the correlation across species between the species means of the set and of the reference, using species with at least three finite values (median and minimum over the 12 measures).}}
\label{stab:anat}
\end{table}

\section{{Supplementary Analyses with Conditional CFM}}
\label{sec:si_historical}
{This section presents supplementary analyses using conditional CFM and the taxonomic proxy (Tables~\ref{stab:historical} and~\ref{stab:loso}). All three studies used the initial conditional CFM and the proxy, with their own generator, checkpoint and evaluation protocols. Their results agree in direction with the matched experiments where both exist, but their values are not directly comparable with Table~\ref{stab:phylo_controls} or main-text Fig.~8.}

\paragraph{{Loss on and off.}} {With one training seed and 100 mixture generations per arm, enabling the phylogenetic loss changed nn-CD from 1.913 to 1.740, MMD-CD from 1.844 to 1.800 and coverage from 0.365 to 0.390, and mean pairwise CD across the mixture fell from 11.616 to 10.557. For the four focal species, the loss lowered species-mean nn-CD from 2.44 to 1.85, MMD-CD from 4.55 to 3.91 (14 of 21 real skulls closer, exact paired sign-flip $p=0.036$) and MorphFID-8 from 8.31 to 6.25, while coverage fell from 0.36 to 0.31, and mixture MorphFID-8 fell from 3.88 to 2.11. The mixture spread statistic includes differences between species, so it does not measure within-species diversity. In the controlled supervision runs of Tables~\ref{stab:phylo_controls} and~\ref{stab:focal_controls}, mixture coverage moved the other way (0.56 without the loss, 0.46--0.52 with it), and only the proxy arm improved the focal species. The focal-species gain agrees in direction with the controlled proxy arm of Table~\ref{stab:focal_controls} (1.81 to 1.47, $p=0.008$), so proxy supervision improved focal-species fidelity in two generators, although neither experiment ties the gain to correct relationships.}

\begin{figure}[t]
\centering
\includegraphics[width=\linewidth]{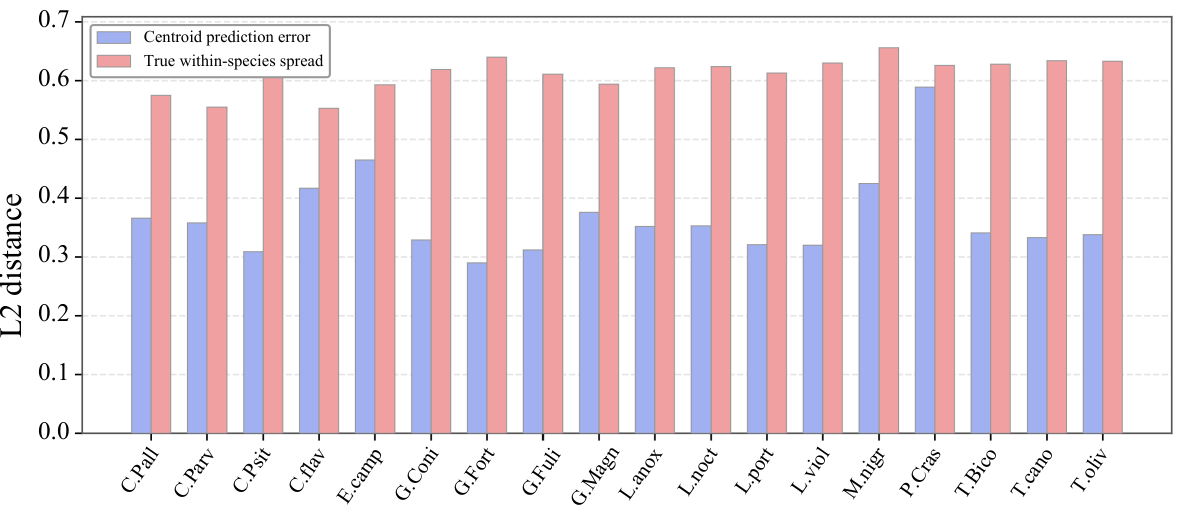}
\caption{{\textbf{Leave-one-species-out centroid interpolation in the trained latent space.} Blue: predicted-to-trained centroid $L_2$ distance. Red: within-species spread. Centroid error is below the observed spread for all 18 species. The representation was not retrained, so each omitted species contributed to training; neighbours follow the implemented proxy (Table~\ref{stab:loso}).}}
\label{sfig:loso}
\end{figure}

\paragraph{{79/21 study.}} {A decoder trained on 79 skulls of 20 species was frozen, and codes were fitted to the 21 withheld skulls. Equal-species mean matched Chamfer distances were 0.5137, 0.5134 and 0.5131 for fitting seeds 42, 43 and 44, against 6.998 for the training-mean code, and clamped SDF errors on 50,000 query points per skull excluded from fitting were 0.00365, 0.00364 and 0.00365, against 0.0252 (normalized units, clamped at 0.1). As in main-text Fig.~8a, this evaluates reconstruction from observed geometry. The same study generated 20 skulls per withheld species with the proxy-conditioned initial CFM (seeds 42--44), pooled Gaussian residuals and empirical residuals, all with centroids predicted from the three nearest proxy neighbours. Equal-species mean nn-CD and MMD-CD were 4.94 and 5.95 for CFM (seed ranges 4.92--4.98 and 5.72--6.16), 4.84 and 5.34 for the pooled Gaussian and 5.18 and 5.33 for empirical residuals, so the initial CFM did not improve on simple controls for unseen species. All 400 attempts produced meshes. The run's runtime configuration was not retained, and we omit its reconstruction MorphFID values because their feature-status fields conflict.}

\begin{table}[t!]
\centering
\footnotesize
\setstretch{1.15}
\setlength{\tabcolsep}{5pt}
{\color{black}
\begin{tabular*}{0.86\linewidth}{@{\extracolsep{\fill}}lrrrr}
\toprule
\multicolumn{5}{l}{\textbf{(a) Loss on and off with the initial CFM: 100 mixture generations per arm}}\\
\addlinespace
Phylogenetic loss & nn-CD & MMD-CD & COV & \shortstack{Mixture pairwise\\CD}\\
\midrule
Off & 1.913 & 1.844 & 0.365 & 11.616\\
On & 1.740 & 1.800 & 0.390 & 10.557\\
\bottomrule
\end{tabular*}
}
\medskip
{{\color{black}
\begin{tabular*}{0.86\linewidth}{@{\extracolsep{\fill}}lrrr}
\toprule
\multicolumn{4}{l}{\textbf{(b) 79/21 study: reconstruction of the 21 withheld skulls}}\\
\addlinespace
Code / fitting seed & Matched CD & \shortstack{Sealed SDF\\MAE} & \shortstack{Extracted/\\attempted}\\
\midrule
Fitted / 42 & 0.5137 & 0.00365 & 21/21\\
Fitted / 43 & 0.5134 & 0.00364 & 21/21\\
Fitted / 44 & 0.5131 & 0.00365 & 21/21\\
Training-mean code & 6.998 & 0.0252 & 21/21\\
\bottomrule
\end{tabular*}
}
\medskip
{\color{black}
\begin{tabular*}{0.86\linewidth}{@{\extracolsep{\fill}}lrrr}
\toprule
\multicolumn{4}{l}{\textbf{(c) 79/21 study: generation for the four withheld species}}\\
\addlinespace
Method / generator seed & nn-CD & MMD-CD & \shortstack{Extracted/\\attempted}\\
\midrule
Initial CFM / 42 & 4.975 & 5.985 & 80/80\\
Initial CFM / 43 & 4.918 & 5.720 & 80/80\\
Initial CFM / 44 & 4.926 & 6.159 & 80/80\\
Initial CFM, mean of seeds & 4.940 & 5.955 & n/a\\
Pooled Gaussian residual & 4.842 & 5.343 & 80/80\\
Empirical residual & 5.178 & 5.329 & 80/80\\
\bottomrule
\end{tabular*}
}}{}
\caption{{Supplementary conditional-CFM analyses under study-specific protocols (Chamfer values $\times10^{-3}$, coverage as a fraction, sealed SDF MAE in normalized SDF units clamped at 0.1). (a)~One training seed. Mixture pairwise CD includes differences between species. (b)~Codes fitted to the withheld skulls' SDF observations with the decoder frozen, with equal weight per withheld species. The sealed SDF queries were not used for fitting. (c)~Equal-species means over the four withheld species after averaging two surface-sampling seeds, with 20 skulls per species for each method and seed. Panels (b) and (c) come from one 79/21 study whose protocol differs from main-text Fig.~8. Attempt counts are n/a (not applicable) for the mean of seeds.}}
\label{stab:historical}
\end{table}

\paragraph{{Leave-one-species-out analysis.}} {For each of the 18 species with at least four scans, we predicted the latent centroid as the inverse-distance-weighted mean of the centroids of its three nearest proxy neighbours and generated skulls around it, without retraining the representation (Table~\ref{stab:loso}, Fig.~\ref{sfig:loso}). Predicted centroids lay on average 0.366 from the trained centroids, against a mean within-species spread of 0.616, and the error was below the spread for every species, most narrowly for \textit{P.~crassirostris} (0.589 against 0.626). The mean row gives the arithmetic mean of the displayed species values. Each held-out species was part of training, and the phylogenetic loss had positioned its codes relative to the other species, so these values describe how the trained latent space arranges species by the proxy rather than prediction for species absent from training. The neighbours include the proxy's genus-parsing error, which pairs \textit{C.~flaveola} (\textit{Coereba}) with \textit{Camarhynchus}. The retrained holdout of main-text Section~5.9 tests prediction directly.}

{The generated-to-real distance ratio is lower for \textit{C.~psittacula} ($1.12\times$) and \textit{G.~fortis} ($1.13\times$) than for \textit{P.~crassirostris} ($1.46\times$), whose centroid prediction error is also the largest (Table~\ref{stab:loso}). This pattern describes interpolation in a representation trained with the implemented proxy; it does not independently validate evolutionary structure. The small species cohorts and the proxy errors limit its interpretation.}

\begin{table}[t]
\centering
  \resizebox{\textwidth}{!}{  
\begin{tabular}{lccccc}
\toprule
\textbf{Species} & $n$ & \textbf{Nearest neighbors} & \textbf{Centroid err.} & \textbf{Within-sp.} & \textbf{Ratio}\\
\midrule
\textit{C.\,pallidus} & 4 & \textit{C. parvulus}, \textit{C.\,psittacula}, \textit{C.\,flaveola} & 0.366 & 0.575 & $1.33\times$\\
\textit{C.\,parvulus} & 5 & \textit{C.\,pallidus}, \textit{C.\,psittacula}, \textit{C.\,flaveola} & 0.358 & 0.555 & $1.36\times$\\
\textit{C.\,psittacula} & 4 & \textit{C.\,pallidus}, \textit{C.\,parvulus}, \textit{C.\,flaveola} & 0.309 & 0.675 & $1.12\times$\\
\textit{C.\,flaveola} & 5 & \textit{C.\,pallidus}, \textit{C.\,parvulus}, \textit{C.\,psittacula} & 0.417 & 0.553 & $1.39\times$\\
\textit{E.\,campestris} & 5 & \textit{L.\,anoxanthus}, \textit{L.\,noctis}, \textit{L.\,portoricensis} & 0.465 & 0.593 & $1.35\times$\\
\textit{G.\,conirostris} & 7 & \textit{G.\,difficilis}, \textit{G.\,fortis}, \textit{G.\,fuliginosa} & 0.329 & 0.619 & $1.24\times$\\
\textit{G.\,fortis} & 5 & \textit{G.\,conirostris}, \textit{G.\,difficilis}, \textit{G.\,fuliginosa} & 0.290 & 0.640 & $1.13\times$\\
\textit{G.\,fuliginosa} & 6 & \textit{G.\,conirostris}, \textit{G.\,difficilis}, \textit{G.\,fortis} & 0.312 & 0.611 & $1.21\times$\\
\textit{G.\,magnirostris} & 5 & \textit{G.\,conirostris}, \textit{G.\,difficilis}, \textit{G.\,fortis} & 0.376 & 0.594 & $1.27\times$\\
\textit{L.\,anoxanthus} & 5 & \textit{L.\,noctis}, \textit{L.\,portoricensis}, \textit{L.\,violacea} & 0.352 & 0.622 & $1.21\times$\\
\textit{L.\,noctis} & 4 & \textit{L.\,anoxanthus}, \textit{L.\,portoricensis}, \textit{L.\,violacea} & 0.353 & 0.624 & $1.23\times$\\
\textit{L.\,portoricensis} & 5 & \textit{L.\,anoxanthus}, \textit{L.\,noctis}, \textit{L.\,violacea} & 0.321 & 0.613 & $1.20\times$\\
\textit{L.\,violacea} & 5 & \textit{L.\,anoxanthus}, \textit{L.\,noctis}, \textit{L.\,portoricensis} & 0.320 & 0.630 & $1.16\times$\\
\textit{M.\,nigra} & 4 & \textit{E.\,campestris}, \textit{L.\,anoxanthus}, \textit{L.\,noctis} & 0.425 & 0.656 & $1.23\times$\\
\textit{P.\,crassirostris} & 4 & \textit{C.\,pallidus}, \textit{C.\,parvulus}, \textit{C.\,psittacula} & 0.589 & 0.626 & $1.46\times$\\
\textit{T.\,bicolor} & 5 & \textit{T.\,canora}, \textit{T.\,olivacea}, \textit{E.\,campestris} & 0.341 & 0.628 & $1.18\times$\\
\textit{T.\,canora} & 5 & \textit{T.\,bicolor}, \textit{T.\,olivacea}, \textit{E.\,campestris} & 0.333 & 0.634 & $1.22\times$\\
\textit{T.\,olivacea} & 4 & \textit{T.\,bicolor}, \textit{T.\,canora}, \textit{E.\,campestris} & 0.338 & 0.633 & $1.19\times$\\
\midrule
\textbf{Mean} & & & \textbf{{0.366}} & \textbf{{0.616}} & $\mathbf{1.25\times}$\\
\bottomrule
\vspace{0.3cm}
\end{tabular}
}
\caption{{Leave-one-species-out centroid interpolation in the trained latent space for the 18 species with at least four scans. The representation was not retrained, so each held-out species contributed to training. Neighbours come from the implemented proxy. Centroid error and within-species spread are latent $L_2$ distances, and the ratio is the generated-to-nearest-real distance divided by the within-species distance for the omitted species (lower is better).}}
\label{stab:loso}
\end{table}

\end{document}